\newif\ifdoublecolumnlook
\doublecolumnlooktrue
\newif\ifnaturelook
\naturelookfalse
\ifdefined\NatureLook\doublecolumnlooktrue\naturelooktrue\fi
\ifdefined\NatureSubmissionLook\doublecolumnlookfalse\naturelookfalse\fi
\newif\ifpralook
\ifdoublecolumnlook\ifnaturelook\else\pralooktrue\fi\fi
\ifdoublecolumnlook
  
  \ifnaturelook  
    \documentclass[10pt,a4paper,twoside]{article}
  \else
    \documentclass[aps,pra,reprint,superscriptaddress]{revtex4-2}
  \fi
\else
    \documentclass[12pt,a4paper]{article}
    
\fi

\ifpralook\else
  \ifdoublecolumnlook
    \usepackage[textwidth=175mm,textheight=253mm,inner=17.5mm,
      bottom=21mm,footskip=8mm]{geometry}
  \else
    \usepackage[margin=1in]{geometry}
  \fi
\fi
\usepackage{setspace}
\ifdoublecolumnlook
  \newcommand{\singlecolwidth}{\columnwidth}   
\else
  \newcommand{\singlecolwidth}{89mm}           
\fi
\ifdoublecolumnlook\else\fi
\ifpralook\else\usepackage{lineno}\fi 
 \usepackage{multirow}
\ifdoublecolumnlook\else\linenumbers\fi
\usepackage{makecell}

\usepackage[T1]{fontenc}
\usepackage[utf8]{inputenc}
\usepackage{newtxtext,newtxmath}   

\usepackage{amsmath}
\usepackage{siunitx}
\usepackage{graphicx}
\usepackage{booktabs}
\ifpralook\else
  \usepackage{caption}
  \ifdoublecolumnlook\else\ifdefined\CaptionLineNumbers
    \DeclareCaptionTextFormat{linenumbered}{\internallinenumbers #1\par}
  \fi\fi
\fi

\ifpralook\else
  \usepackage{authblk}

\fi

\ifpralook
\else
  \usepackage[super,sort&compress,numbers]{natbib}
\fi

\usepackage[hidelinks]{hyperref}
\usepackage{orcidlink}
\usepackage{comment}

\ifdoublecolumnlook\else\ifdefined\FiguresAtEnd
  \usepackage[nolists,nomarkers,noheads,figuresonly]{endfloat}
\fi\fi

\newenvironment{summaryparagraph}{%
  \ifpralook
    \begin{abstract}
  \else
    \begingroup\bfseries
  \fi
}{%
  \ifpralook
    \end{abstract}\maketitle
  \else
    \par\endgroup\vspace{1em}
  \fi
}

\newcommand{\extendeddatafigures}{%
  \setcounter{figure}{0}%
  \gdef\thefigure{\arabic{figure}}%
  \gdef\figurename{Extended Data Fig.}%
  \gdef\tablename{Extended Data Table}%
}

\ifdoublecolumnlook
    \ifnaturelook
      \input{nature-look.tex}
    \else

        \RenewDocumentEnvironment{table}{O{t}}{\begin{table*}[t]}{\end{table*}}
        
    \fi
\fi

\begin{document}

\title{Continuous superradiant lasing on a narrow optical transition}

\newcommand{\affilAmsterdam}{Van der Waals-Zeeman Institute, Institute of Physics,
University of Amsterdam, Amsterdam, The Netherlands}
\newcommand{\affilQuSoft}{QuSoft, Science Park 123, 1098 XG Amsterdam, The Netherlands}
\newcommand{\affilBonn}{Physikalisches Institut, University of Bonn, Nussallee 12, 53115 Bonn, Germany}
\newcommand{\affilINRiM}{Istituto Nazionale di Ricerca Metrologica (INRiM), Turin, Italy}
\newcommand{\affilCopenhagen}{NNF Quantum Computing Programme, Niels Bohr Institute,
University of Copenhagen, Copenhagen, Denmark}
\newcommand{\correspondencenote}{Correspondence should be addressed to
  \href{mailto:contSR@strontiumBEC.com}{contSR@strontiumBEC.com}.}

\ifpralook
\author{Benedikt Heizenreder\,\orcidlink{0000-0001-5611-4144}}
\affiliation{\affilAmsterdam}
\affiliation{\affilQuSoft}

\author{Simon B. J\"{a}ger\,\orcidlink{0000-0002-2585-5246}}
\affiliation{\affilBonn}

\author{Francesca Famà\,\orcidlink{0009-0009-3162-8806}}
\affiliation{\affilAmsterdam}
\affiliation{\affilQuSoft}
\affiliation{\affilINRiM}

\author{Camila Beli Silva\,\orcidlink{0000-0002-3862-1203}}
\affiliation{\affilAmsterdam}
\affiliation{\affilQuSoft}
\affiliation{\affilCopenhagen}

\author{Ying Cui\,\orcidlink{0009-0001-9138-7962}}
\affiliation{\affilAmsterdam}
\affiliation{\affilQuSoft}

\author{Stefan A. Sch\"{a}ffer\,\orcidlink{0000-0002-5296-0332}}
\affiliation{\affilAmsterdam}
\affiliation{\affilQuSoft}
\affiliation{\affilCopenhagen}

\author{Ananya Sitaram\,\orcidlink{0000-0002-6548-4515}}
\affiliation{\affilAmsterdam}
\affiliation{\affilQuSoft}

\author{Florian Schreck\,\orcidlink{0000-0001-8225-8803}}
\thanks{\correspondencenote}
\affiliation{\affilAmsterdam}
\affiliation{\affilQuSoft}

\else
\author[1,2]{Benedikt Heizenreder\,\orcidlink{0000-0001-5611-4144}}

\author[3]{Simon B. J\"{a}ger\,\orcidlink{0000-0002-2585-5246}}

\author[1,2,4]{Francesca Famà\,\orcidlink{0009-0009-3162-8806}}

\author[1,2,5]{Camila Beli Silva\,\orcidlink{0000-0002-3862-1203}}

\author[1,2]{Ying Cui\,\orcidlink{0009-0001-9138-7962}}

\author[1,2,5]{Stefan A. Sch\"{a}ffer\,\orcidlink{0000-0002-5296-0332}}

\author[1,2]{Ananya Sitaram\,\orcidlink{0000-0002-6548-4515}}

\author[1,2]{Florian Schreck\,\orcidlink{0000-0001-8225-8803}%
  \thanks{\correspondencenote}}

\affil[1]{\affilAmsterdam}
\affil[2]{\affilQuSoft}
\affil[3]{\affilBonn}
\affil[4]{\affilINRiM}
\affil[5]{\affilCopenhagen}
\fi

\ifpralook\date{\today}\else\date{}\fi
\ifdoublecolumnlook\else\maketitle\fi

\begin{summaryparagraph}
Lasers with ultrastable frequency are at the foundation of optical clocks~\cite{fortierOpticalAtomicClocks2026,Aeppli_Atomic_Clock_Frequency_2026,liuZeroDeadTimeStrontiumLattice2025,Arnold2026}, which enable precision tests of fundamental physics~\cite{kennedyPrecisionMetrologyMeets2020,arakawaProbingUltralightDark2026,chuExploringDynamicalInterplay2025,takamotoTestGeneralRelativity2020,bothwellResolvingGravitationalRedshift2022} and relativistic geodesy at the centimetre scale~\cite{Mehlstaubler2018}. Such lasers inherit their frequency stability from the length stability of a sophisticated optical resonator, which is hard to maintain outside a well-controlled laboratory environment and limited by thermal and technical noise. 
Superradiant lasers promise to overcome these limitations by storing their coherence in the atoms making up the laser's gain medium~\cite{Prospects_for_a_Millihertz_Linewidth_Meiser,Chen09,Rugged_mHz_Linewidth_Haonan_2020}. Superradiant lasing has been demonstrated on narrow optical atomic transitions with pulsed and quasi-continuous operation~\cite{bohnetSteadystateSuperradiantLaser2012a, norciaSuperradianceMillihertzLinewidth2016, Schaffer2020, Laske2019, norciaColdStrontiumLaserSuperradiant2016, Kristensen2023}. However, atom loss has prevented truly continuous operation, which severely limits their suitability for clocks. 
Here we demonstrate a superradiant laser on a narrow optical transition in \textsuperscript{88}Sr that operates truly continuously. We sustain emission for more than eight hours and measure a linewidth of 320(40)\,Hz, more than 20 times below the natural linewidth. The emission frequency remains anchored to the atomic transition, making the laser itself an atomic frequency reference. 
Our results provide a route towards a laser that reaches short-term frequency stability comparable to that of the best resonator-stabilized lasers. Furthermore, truly continuous operation gives us access to physics beyond the transient regime, enabling the exploration of robust nonequilibrium steady states and sustained coherent collective dynamics.
\end{summaryparagraph}

\begin{figure*}[t]
\centering
\includegraphics[width=0.9 \textwidth]{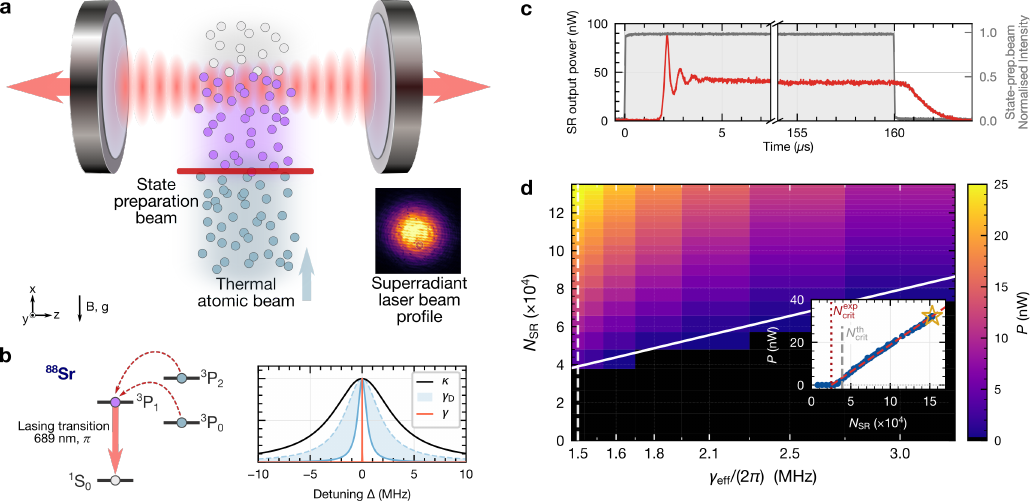}
\caption{
\textbf{Continuous superradiant lasing on a narrow optical transition.}
\textbf{(a)} Schematic of the experimental apparatus.
A thermal beam of \textsuperscript{88}Sr atoms is transversely cooled and the atoms are prepared in long-lived ${^3\!P_{0,2}}$ states (blue dots). A state preparation beam optically pumps atoms to the upper lasing state, ${^3\!P_1}$ (purple dots). These atoms superradiantly lase while traversing a cavity. The photo inset shows the intensity profile of the emitted laser beam.
\textbf{(b)} (left) Level diagram of \textsuperscript{88}Sr showing the ${^3\!P_1} \to {^1\!S_0}$ lasing transition at 689\,nm and the action of the state preparation beam. (right) Spectrum of the atomic transition labelled by its linewidth $\gamma$, the Doppler broadened transition for the accessible range of linewidths $\gamma_\mathrm{D}$ and the bare cavity with linewidth $\kappa$.
\textbf{(c)} Continuous superradiant lasing (red line, left axis). Shortly after switch-on of the state preparation beam (grey line, right axis; shaded region: beam on), lasing starts with relaxation oscillations, settles into steady state, and persists until the preparation beam is switched off.
\textbf{(d)} Laser output power as a function of effective linewidth $\gamma_\mathrm{eff}\sim\gamma_\mathrm{D}$ and intracavity ${^3\!P_1}$ atom number $N_\mathrm{SR}$. The theoretical lasing threshold [Eq.~\eqref{eq:NC}] is shown as a white line.
Inset: output power versus $N_\mathrm{SR}$ at our reference $\gamma_\mathrm{eff}/(2\pi) = 1.47(4)$\,MHz, white dashed line in main graph. Vertical dashed lines indicate the lasing onset obtained from the fit, $N^{\rm exp}_\mathrm{crit}$, and theory, $N^{\rm th}_\mathrm{crit}$. Red dashed line: linear fit to the points above 1\,nW. Star: reference operating point.
Error bars indicate the standard deviation of the shot-averaged signals across repeated measurements at each scan point.}
\label{fig:Figure_1}
\end{figure*}

\ifdoublecolumnlook\else
  \ifdefined\FiguresAtEnd
    \vspace{2cm}
  \else
    \clearpage
  \fi
\fi

\noindent 
Optical atomic clocks are the most precise measurement devices ever built, reaching fractional frequency uncertainties below $10^{-18}$~\cite{fortierOpticalAtomicClocks2026,Aeppli_Atomic_Clock_Frequency_2026,liuZeroDeadTimeStrontiumLattice2025,Arnold2026}. This precision originates from the ability to transfer the frequency of an ultranarrow optical atomic transition onto a laser, making it a frequency standard. Current optical clocks are \emph{passive} frequency standards, in which the laser interrogates atoms spectroscopically. 
To ensure a narrow spectroscopy signal, 
the laser is frequency stabilized on short timescales to the mode of an ultrastable optical resonator (cavity). However, even small mechanical or thermal fluctuations of the cavity length are mapped directly onto the laser frequency, limiting the short-term stability of the clock~\cite{parkeLaserFractionalFrequency2026,Lee_Frequency_Stability_2026}.

An elegant alternative to such \emph{passive} frequency standards is an \emph{active} one, in which the atoms themselves emit light at the frequency of the clock transition. 
Hydrogen masers have demonstrated this possibility for decades in the microwave domain~\cite{daiReviewDevelopmentHydrogen2022}. 
In the optical domain, an active frequency reference can be realized by operating in the ``bad-cavity regime''.
In contrast to the conventional lasing regime~\cite{schawlowInfraredOpticalMasers1958}, the lasing cavity linewidth in the bad-cavity regime greatly exceeds the gain medium linewidth and the laser's frequency is mostly determined by the gain medium. Continuous-wave bad-cavity lasing has been demonstrated with gain medium linewidths from one to several hundred megahertz~\cite{kuppensQuantumLimitedLinewidthBadCavity1994,vanexterTheoryLinewidthBadcavity1995,zhangExtremelyBadcavityLaser2024}, and the influence of the bad-cavity regime on laser properties such as the linewidth was explored. 
However, these gain media are too broad and too strongly perturbed to serve as accurate atomic frequency references. 
Instead, a narrow atomic transition can provide the gain, allowing the laser itself to be an accurate atomic frequency reference. In this case, the atomic dipoles radiate collectively (superradiantly), which enhances the emission rate far beyond that of independently emitting atoms and enables lasing despite the weakness of the narrow transition~\cite{Dicke_Coherence_Spontaneous_Radiation_1954,ferioliNonequilibriumSuperradiantPhase2023}. Such superradiant (SR) lasers are predicted to be able to reach linewidths down to the millihertz level~\cite{Chen09,Prospects_for_a_Millihertz_Linewidth_Meiser}.

First steps towards realizing such a device have been taken, with pulsed and quasi-continuous superradiant emission of less than one second duration on narrow optical atomic transitions and engineered Raman transitions confirming the collective emission physics~\cite{norciaSuperradianceMillihertzLinewidth2016, Schaffer2020, Laske2019, norciaColdStrontiumLaserSuperradiant2016, Kristensen2023, bohnetSteadystateSuperradiantLaser2012a}. 
Collective emission has also been studied in hot atomic beams, where phase coherence was imprinted with a phase mask~\cite{kimCoherentSingleatomSuperradiance2018}.
However, a truly continuous superradiant laser, whose linewidth is not limited by the finite pulse duration, and that enables access to collective dynamics beyond that of a conventional laser, has remained an outstanding goal.

Here we demonstrate a truly continuous superradiant laser on the $^1\!S_0 \to {^3\!P_1}$ intercombination line of $^{88}$Sr at 689\,nm, with transition linewidth $\gamma/(2\pi)\,=\,7.48\,$kHz, operating in the bad-cavity regime.
In our experiment, a thermal atomic beam continuously replenishes excited-state atoms coupled to the cavity mode, and superradiant lasing persists. We realize an operation time of more than eight hours without interruption, with a steady output power of $65(1)$\,nW. We measure a linewidth of $320(40)$\,Hz, an upper bound on the intrinsic superradiant laser linewidth, more than 20 times below the natural linewidth of the atomic transition and more than 3\,000 times below the Doppler broadening, which is the main decoherence mechanism. The frequency stays anchored to the atomic transition with an uncertainty of a few kHz. We compare the experimental observations of critical atom number, cavity pulling, and linewidth scaling to a simple superradiant-laser theory and find good agreement. 
In addition, by exploring the boundaries of the metrologically useful regime of this superradiant laser, we uncover signatures of multi-component superradiance, a dynamical regime characterized by long-lived, coherent amplitude oscillations.

\begin{figure*}[t]
    \centering
    \includegraphics[width=\textwidth]{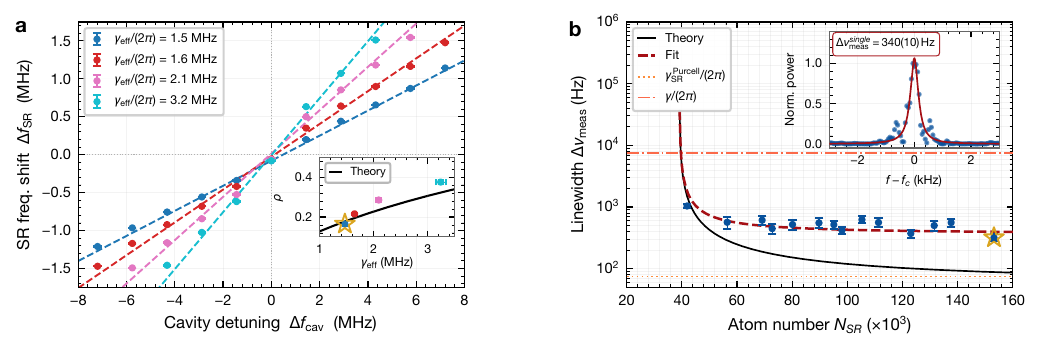}
    \caption{
        \textbf{Bad-cavity operation and superradiant linewidth.}
        \textbf{(a)}~Cavity pulling in the bad-cavity regime.
        The superradiance frequency shift $\Delta f_\mathrm{SR}$ is measured as a function of the cavity detuning $\Delta f_\mathrm{cav}$, for four separate Doppler linewidths $\gamma_\mathrm{D}\sim\gamma_\mathrm{eff}$. Each point is the mean of five measurements. Dashed lines are linear fits to the individual scans whose slope is the cavity-pulling coefficient $\rho(\gamma_\mathrm{eff})$.  Inset: $\rho$ versus $\gamma_\mathrm{eff}$. The solid line is the theoretical prediction, Eq.~\eqref{eq:rho}.  
        \textbf{(b)}~Superradiant emission linewidth $\Delta\nu_\mathrm{meas}$ as a function of intracavity ${^3\!P_1}$ atom number $N_\mathrm{SR}$, measured at $\gamma_\mathrm{eff}/(2\pi) = 1.47(4)$\,MHz (blue disks, mean of 25-30 measurements), reaching a minimum linewidth of $320(40)$\,Hz at our reference operating point (star). The solid black line shows the analytical result for the linewidth Eq.~\eqref{eq:linewidth_main}, with a minimum of $\gamma_\mathrm{SR}/(2\pi)\approx90$\,Hz over the observed range. The dashed red line is the analytical result including an offset for the finite linewidth of the probe laser and technical noise contributions in the SR cavity. 
        The natural linewidth of the $^1\!S_0 \to {^3\!P_1}$ transition is shown as an orange dash-dotted line, and the Purcell-limited linewidth of $\gamma_\mathrm{SR}^{\,\mathrm{Purcell}}/(2\pi)=75$\,Hz is shown as an orange dotted line.
        Inset: single-shot beat-note spectrum for $N_\mathrm{SR}=1.53(2) \times 10^5$ atoms (blue dots) fitted with a Lorentzian (red line).
        Stars indicate our reference operating point. Error bars are the standard error about the mean from measurement scatter combined in quadrature with propagated Lorentzian fit uncertainties for individual shots. 
    }
    \label{fig:figure2}
\end{figure*}

\subsection*{Superradiant lasing from a beam of cold atoms}

We demonstrate continuous superradiant lasing on an optical transition in the bad-cavity regime using a thermal beam of excited $^{88}$Sr atoms passing through the mode of a linear optical cavity~(Fig.~\ref{fig:Figure_1}a).
Atoms couple via the 7.48\,kHz-wide $^1\!S_0 \to {}^3\!P_1$ intercombination line at 689\,nm to the resonant cavity with decay rate $\kappa/(2\pi) = 6.79(2)$\,MHz. 
Before entering the cavity, the atoms are cooled and shelved in the metastable $^3\!P_{0,2}$ states using Swept Velocity Shelving (SVS)~\cite{famaCollimationDenseAtomic2026}. 
The cooling method delivers a state-heralded supply of collimated atoms that can be transferred to our chosen lasing state ($^3\!P_1$) with a state-preparation laser just before the cavity~(Fig.~\ref{fig:Figure_1}b, left). As the atomic beam traverses the cavity mode, excited state atoms are continuously introduced, maintaining the inverted gain medium and sustaining lasing. 

The number of atoms in the excited lasing state, $N_\mathrm{SR}$, is controlled via the intensity of the state-preparation beam and reaches $N_\mathrm{SR} = 1.57(8) \times 10^5$ at a maximum flux of $ 3.6(1)\times10^{11}$\,atoms\,s$^{-1}$ through the cavity mode. Atoms remaining in the metastable states do not couple directly to the cavity mode and therefore do not influence lasing. The transit time of the atoms through the cavity mode is $\approx$ 0.45\,$\mu$s, which gives rise to a transit time broadening of $\gamma_\mathrm{T}/(2\pi) \approx 360$\,kHz.
The transit time is 50 times shorter than the natural lifetime of the lasing state, which makes collective enhancement of the emission rate necessary for lasing.
The dominant decoherence mechanism in the system is Doppler broadening of the atomic transition when interacting with light in the cavity. We precisely control the Doppler broadening by adjusting experimental parameters during the SVS stage and can reduce it to a minimum of $\gamma_\mathrm{D}/(2\pi)\approx1$\,MHz, while still maintaining the necessary excited state atom number to sustain lasing.
The effective coherence decay rate can thus be expressed by taking into account both the transit time and the Doppler broadening: $\gamma_\mathrm{eff}=\gamma_\mathrm{D}+\gamma_\mathrm{T}+\gamma$, where in our case $\gamma_\mathrm{eff} \sim \gamma_\mathrm{D} \gg \gamma_\mathrm{T}, \gamma$. 
The cavity decay rate $\kappa$ exceeds $\gamma_\mathrm{eff}$, placing the system in the bad-cavity regime~(Fig.~\ref{fig:Figure_1}b, right).
\\
The onset of superradiant lasing occurs when the collective atomic gain overcomes the decoherence mechanisms.
We initiate lasing by switching on the state-preparation beam, populating the lasing state, and allowing the atoms to couple to the cavity.
At the onset of lasing, we observe relaxation oscillations lasting $\approx$ 5\,$\mu$s before the emission settles into steady-state superradiant lasing, as shown in Fig.~\ref{fig:Figure_1}c.
Steady-state lasing persists until the state-preparation beam is switched off.
To extract the lasing threshold atom number, we measure the lasing output power in the $(\gamma_\mathrm{D},N_\mathrm{SR})$ plane, see Fig.~\ref{fig:Figure_1}d. 
We observe reasonable agreement with the theoretically predicted threshold (see SI), indicated by the white line, and given by  
    \begin{equation}
        N_\mathrm{crit} = \frac{\kappa\,\gamma_\mathrm{eff}}{2g_0^2},
        \label{eq:NC}
    \end{equation}
where $g_0/(2\pi) = 11.3$\,kHz is the single-atom cavity coupling rate. 
Above threshold, we observe a linear power scaling with $N_\mathrm{SR}$, which is characteristic of continuous superradiant lasing in steady-state, when energy conservation prevents the quadratic scaling observed for transient superradiant bursts~\cite{tangProspectsSuperradiantLaser2022}. 
At our reference operating point (marked in the inset by a star, $N_\mathrm{SR} \approx 1.53(2) \times 10^5$, $\gamma_\mathrm{eff}/(2\pi) \approx 1.47(4)$\,MHz), the total emitted power is $65(1)$\,nW.

\begin{figure*}[t]
\centering
\includegraphics[width= \textwidth]{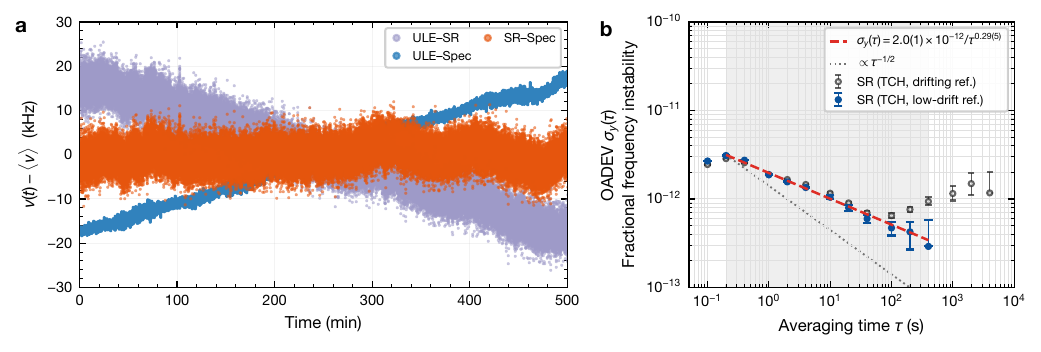}
\caption{
\textbf{Uninterrupted superradiant lasing and long-term stability.}
\textbf{(a)} Raw beat-note frequency measurements over 500 min of continuous SR laser operation at $\nu_0 = 434.829$~THz ($100\,\mathrm{ms}$ gate time). 
We compare the superradiant (SR), spectroscopy- (Spec) and ULE-referenced (ULE) lasers in a three-cornered hat configuration. SR--Spec (orange), ULE--Spec (blue), and ULE--SR (purple) beats are shown with their mean value subtracted.
\textbf{(b)} The fractional frequency instability $\sigma_y(\tau)$ of the SR laser extracted via the overlapping Allan deviation of the three-cornered hat (TCH) measurement. Grey circles: 500~min dataset (same as \textbf{a}), where ULE laser drift ($4215 (1)\,\mathrm{Hz/h}$) dominates the variance decomposition at $\tau \gtrsim 100$~s. Blue disks: independent 140~min dataset recorded with lower ULE drift ($290 (10)\,\mathrm{Hz/h}$).
A power-law fit over the grey shaded region yields a temporal scaling of $\tau^{-0.29(5)}$.
The instability reaches $2.9 \times 10^{-13}$ at $\tau = 400$~s. The dotted line shows the ideal phase diffusion scaling $\sigma_y \propto \tau^{-1/2}$.
Error bars indicate $1\sigma$ confidence intervals, calculated using a chi-squared distribution with conservative $T/\tau - 1$ degrees of freedom where $\tau$ is the cluster time~\cite{Riley2008}.
}
\label{fig:Figure_3}
\end{figure*}

\subsection*{Bad-cavity operation and superradiant linewidth}

Because superradiant lasers exploit the phase coherence of  atomic evolution, temporal drift and noise in the resonance condition of the fast-decaying cavity are suppressed in the emitted frequency spectrum. This suppression of the normal cavity pulling behaviour seen in lasers can be expressed by
\begin{equation}
    \rho = \frac{\Delta f_\mathrm{SR}}{\Delta f_\mathrm{cav}}
         = \frac{\gamma_\mathrm{eff}}{\gamma_\mathrm{eff} + \kappa} < 1,
    \label{eq:rho}
\end{equation}
where $\Delta f_\mathrm{SR}$ is the shift of the SR output frequency in response to a cavity detuning $\Delta f_\mathrm{cav}$.
In Fig.~\ref{fig:figure2}a we demonstrate a minimum pulling coefficient of $\rho = 0.165(3)$, measured at the lowest effective broadening currently accessible in our apparatus, $\gamma_\mathrm{eff}/(2\pi) = 1.47(4)$\,MHz, confirming operation in the bad-cavity regime.
Since $\gamma_\mathrm{eff}$ is dominated by the Doppler broadening $\gamma_\mathrm{D}$, the cavity pulling $\rho$ can be varied by tuning this parameter.
This behaviour can be seen in the inset of Fig.~\ref{fig:figure2}a, where we show $\rho$ for four values of $\gamma_\mathrm{D}$ (and thus $\gamma_\mathrm{eff}$), together with Eq.~\eqref{eq:rho}, confirming that this simple model already provides a good description of our system.

We next investigate the intrinsic linewidth of the SR laser and its scaling with the intracavity atom number $N_\mathrm{SR}$.
To extract the linewidth of the SR laser, we heterodyne the SR output against an ultra-low expansion (ULE)-cavity-stabilized reference laser (ULE laser).
We measure a minimum linewidth of $\Delta\nu_\mathrm{meas}=320(40)$\,Hz, more than $20$ times below the natural linewidth $\gamma/(2\pi) = 7.48$\,kHz and more than $3\,000$ times below $\gamma_\mathrm{D}$, demonstrating the strong phase coherence of the collective dipole and providing an upper bound for the intrinsic linewidth.
We also find that the linewidth decreases with increasing intracavity atom number, which is consistent with theoretical predictions~\cite{SR_thermal_atomic_beam_Jager_Simon}.
To analyse the scaling quantitatively, we consider a simple analytical model that takes into account transit time and Doppler broadening, both present in our system.
The theoretical estimate for the linewidth is
\begin{equation}
        \gamma_\mathrm{SR} = \frac{\kappa}{4N_\nu}\rho^2 \left( 1+\frac{2g_0^2N_\mathrm{SR}}{\kappa\gamma_\mathrm{eff}} \right),
        \label{eq:linewidth_main}
\end{equation}
where $N_\nu\propto N_\mathrm{SR}$ is the number of intracavity photons (see SI). This expression contains two independent noise contributions, obtained by adding stochastic noise sources to the linewidth equations. The first arises from cavity shot noise, analogous to the Schawlow-Townes mechanism, but suppressed in the bad-cavity regime by the factor $\rho^2$, yielding a contribution proportional to $\kappa\rho^2/(4N_\nu)$. The second arises from atomic injection noise: each incoming atom contributes quantum phase noise to the collective dipole. This contribution scales as $g_0^2\rho^2N_\mathrm{SR}/(2\gamma_\mathrm{eff}N_\nu)$. 
The analytical model (solid black line in Fig.~\ref{fig:figure2}b) predicts a minimum intrinsic linewidth of $\gamma_\mathrm{SR}/(2\pi)\approx90$\,Hz at the lowest measured broadening, $\gamma_\mathrm{eff}/(2\pi) = 1.47(4)$\,MHz, and an estimated intracavity atom number of $N_\mathrm{SR} = 1.53(2) \times 10^5$. The lowest measured linewidth lies above the theoretically predicted minimum, which can be attributed to technical noise sources that are not captured by the idealized analytical model. 
If we introduce an additional atom-independent frequency noise contribution originating from technical-noise-induced cavity pulling or the finite linewidth of the ULE laser, $\Delta\nu_\mathrm{add}=310(60)$\,Hz (see Methods), we find good agreement between the data and the analytical model (red dashed line), indicating that the measured linewidth could be reduced even further by addressing these noise contributions.
Nevertheless, the narrow measured linewidth compared to the natural linewidth of the atomic transition already demonstrates the exceptional spectral purity of the superradiant laser.

\begin{figure*}[t]
\centering
\includegraphics[width= \textwidth]{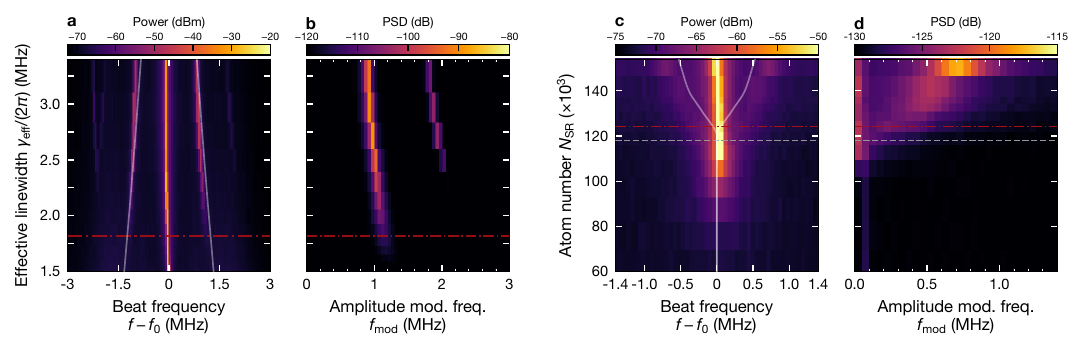}
\caption{
\textbf{Single-frequency to multi-component lasing.}
\textbf{(a)} Beat-note frequency spectrum of the SR laser against the ULE laser as a function of effective linewidth $\gamma_\mathrm{eff}$. The frequency axis is referenced to the single-mode SR peak at the lowest $\gamma_\mathrm{eff}$. Sidebands emerge above $\gamma_\mathrm{eff}^\mathrm{MCSR}/(2\pi)~=~1.8(1)$\,MHz (horizontal red dash-dotted line). White solid lines: theoretically predicted sideband frequencies.
\textbf{(b)} Intensity noise PSD recorded on an avalanche photodiode
(DC component subtracted), same scan range as in~(a). A non-zero intensity modulation emerges, consistent with (a), above the same $\gamma_\mathrm{eff}^\mathrm{MCSR}/(2\pi)$. 
\textbf{(c)} Beat-note spectrum as a function of intracavity atom number $N_{\rm SR}$ for $\gamma_\mathrm{eff}/(2\pi) = 4.5(1)$\,MHz. 
A significant lasing peak emerges around $N_\mathrm{crit} = 1.13(1)\times10^5$ atoms and sidebands develop above $N_\mathrm{MCSR} = 1.24(5)\times10^5$ (red dash-dotted line).
The theoretical critical atom number $N_\mathrm{crit}^\mathrm{theo} = 1.18\times10^5$ is shown by the horizontal grey dashed line. White solid lines: theoretically predicted sideband frequencies.
\textbf{(d)} Intensity noise PSD for the same scan range as in (c). A non-zero intensity modulation consistent with (c) emerges above $N_\mathrm{MCSR}$. 
}
\label{fig:Figure_4}
\end{figure*}

\subsection*{Long-term stability}

For stable long-term operation of the laser, the cavity resonance frequency must be kept close to the atomic transition. We achieve this through a self-locking scheme: we stabilize the cavity to a frequency derived from the superradiant output itself, forcing the cavity length to track the superradiant emission frequency (see Methods). This scheme, enabled by the suppressed cavity pulling, renders the superradiant laser a long-term stable atomic frequency reference.

To demonstrate uninterrupted operation and investigate long-term stability, we continuously monitor the SR output frequency for $500$\,min (Fig.~\ref{fig:Figure_3}a). 
The measurement is performed in a three-cornered-hat configuration~\cite{vernotteThreecorneredHatAllan2016} so that the individual noise contributions of three independent lasers can be determined.
Of the two lasers we compare the SR laser to, one is stabilized to the $^1\!S_0\to{}^3\!P_1$ transition in an atomic vapour cell, providing a long-term atomic frequency reference at the kHz level (Spec laser), and the other is locked to a ULE cavity, but drifts with $4215 (1)\,\mathrm{Hz/h}$ over the span of the measurement (ULE laser).
While the ULE--Spec and ULE--SR beat frequencies exhibit a clear drift over the full $500$\,min interval, the SR--Spec beat remains nearly constant, with a standard deviation of 2.3\,kHz from the mean, demonstrating that the SR frequency is self-referenced to the collective atomic transition frequency.

We further use the three-cornered-hat analysis to determine the fractional frequency instability of the SR laser itself (Fig.~\ref{fig:Figure_3}b). For the $500$\,min dataset, the long-term stability is limited by the strong drift of the ULE laser. 
Thus, we analyse an independent $140$\,min dataset recorded for a reduced ULE laser drift of $290(10)\,\mathrm{Hz/h}$. 
This measurement yields a fractional Allan deviation of $\sigma_y(1\,\mathrm{s}) = 2.0\times10^{-12}$ and reaches $\sigma_y = 2.9\times10^{-13}$ at $\tau = 400$\,s. The observed scaling is shallower than the ideal phase-diffusion limit $\sigma_y \propto \tau^{-1/2}$, consistent with technical noise present in our system (see Methods and SI)~\cite{kazakovUltimateStabilityActive2022}.

\subsection*{Multi-component superradiance}

Having established our setup as a superradiant laser with a narrow optical emission frequency suitable as a frequency reference, we now push the experiment beyond the limits of this metrological regime. By increasing the Doppler linewidth, we deliberately drive the system away from steady-state superradiant lasing. As $\gamma_\mathrm{eff}$ increases, the mean output power decreases while the measured linewidth broadens, in overall agreement with the analytical model of Eq.~\eqref{eq:linewidth_main} (Ext. Data Fig.~\ref{fig:Figure_6_methods}). Remarkably, however, the deterioration of metrological performance is accompanied by the emergence of a new dynamical regime: we observe characteristic signatures of multi-component superradiance (MCSR), namely long-lived collective amplitude oscillations~\cite{SR_thermal_atomic_beam_Jager_Simon}.
	
While such instabilities ultimately limit the regime of stable frequency-reference operation, they have long been of interest in bad-cavity lasers as a route to rich nonlinear dynamics, including period doubling and chaos~\cite{abrahamOverviewInstabilitiesLaser1985,ogawaQuasiperiodicInstabilityChaos1988}. Their controlled observation in our experiment therefore demonstrates the versatility of our platform and enables the exploration of novel nonlinear dynamics in superradiant lasers. 
	
While scanning the effective Doppler broadening $\gamma_\mathrm{eff}$, we observe above $\gamma_\mathrm{eff}^\mathrm{MCSR}/(2\pi) = 1.8(1)$\,MHz the appearance of sidebands in the heterodyne beat spectrum between the SR laser and the ULE laser (Fig.~\ref{fig:Figure_4}a). At $\gamma_\mathrm{eff}^\mathrm{MCSR}/(2\pi) = 3.0(1)$\,MHz, the dominant sideband lies at $f_\mathrm{mod,1}^\mathrm{exp} = 0.94(4)$\,MHz and a second harmonic is weakly visible at $f_\mathrm{mod,2}^\mathrm{exp} = 1.86(4)$\,MHz. The appearance of these sidebands indicates a coherent modulation of the emission, which we identify as the MCSR regime.
	
Theory predicts that the observed modulation originates from oscillations of the amplitude mode, which is decoupled from the phase mode of the superradiant laser. We test this prediction in two complementary ways. First, we compare the theoretically predicted amplitude-mode frequency quantitatively with the experimentally measured value. For this, we develop a microscopic description of the superradiant laser based on the atomic beam and calculate the amplitude-resonance frequencies (see Supplementary Information). For the average atom number of the $\gamma_{\rm eff}$ scan, $N_{\mathrm{SR}}=1.57(8) \times 10^5$, the theoretically calculated amplitude-mode frequencies are shown as white lines in Fig.~\ref{fig:Figure_4}a, closely following the measured resonance frequency. At $\gamma_\mathrm{eff}^\mathrm{MCSR}/(2\pi) = 3.0(1)$\,MHz theory predicts $f_\mathrm{mod}^\mathrm{theo} = 0.93(1)$\,MHz. 

Second, we directly probe the amplitude dynamics experimentally. While the heterodyne measurement contains contributions from both the phase and amplitude modes, the amplitude can be accessed independently through the intensity-noise power spectral density (PSD). Figure~\ref{fig:Figure_4}b shows the intensity PSD over the same range of $\gamma_\mathrm{eff}$, with the DC component removed. Peaks emerge above the same threshold, directly demonstrating coherent oscillations of the laser-field amplitude. The peaks cross $f_\mathrm{mod,1}^\mathrm{exp} = 0.93(3)$\,MHz and $f_\mathrm{mod,2}^\mathrm{exp} = 1.86(3)$\,MHz at $\gamma_\mathrm{eff}^\mathrm{MCSR}/(2\pi) = 3.0(1)$\,MHz, consistent with Fig.~\ref{fig:Figure_4}a.  

The rich dynamical behaviour is further highlighted by entering the MCSR regime through a different parameter scan: increasing $N_{\rm SR}$ while fixing $\gamma_\mathrm{eff}/(2\pi) = 4.5(1)$\,MHz~(Fig.~\ref{fig:Figure_4}c, d). Beyond the theoretical lasing transition (grey dashed line) and the onset of MCSR (red dash-dotted line), two sidebands emerge from the carrier, in contrast to the onset at finite frequency observed when varying $\gamma_\mathrm{eff}$. 
At $N_\mathrm{SR} = 1.54\times10^5$ the observed sideband frequency in the beat spectrum ($f_\mathrm{mod}^\mathrm{exp} = 0.70(3)$\,MHz) and the PSD ($f_\mathrm{mod}^\mathrm{exp} = 0.71(2)$\,MHz), show good qualitative agreement with the theoretical prediction $f_\mathrm{mod}^\mathrm{theo} = 0.53(1)$\,MHz.
Together, the agreement between the heterodyne and intensity measurements and the microscopic model provides strong evidence that the observed sidebands originate from the collective amplitude dynamics of MCSR, giving direct access to a regime of nonlinear collective dynamics~\cite{SR_thermal_atomic_beam_Jager_Simon,LaburtheTolra2026}.

\section*{Conclusion}
We have realized a continuous superradiant laser on the $^1\!S_0 \rightarrow {}^3\!P_1$ transition of $^{88}$Sr in a hot-beam architecture~\cite{Rugged_mHz_Linewidth_Haonan_2020}. We measured an upper bound of 320(40)\,Hz on the intrinsic linewidth, which is more than $3\,000$ times narrower than the dominant decoherence rate in the system and more than $20$ times below the natural linewidth of the transition. We confirmed the predicted lasing threshold and cavity-pulling factor and achieved an output power of $65(1)$\,nW. By using the SR laser frequency itself to stabilize the cavity length, we demonstrated uninterrupted operation for over eight hours and extracted a fractional frequency instability of $2.9\times10^{-13}$ after 400~s of averaging using the three-cornered-hat method. Our observations agree reasonably well with a simple superradiant-laser theory~\cite{Prospects_for_a_Millihertz_Linewidth_Meiser,Rugged_mHz_Linewidth_Haonan_2020,SR_thermal_atomic_beam_Jager_Simon,kazakovUltimateStabilityActive2022}.

Continuous SR lasing on the $^{88}$Sr intercombination line is directly transferable to other species with comparably narrow transitions, including $^{40}$Ca and $^{174}$Yb, using the same beam geometry~\cite{Rugged_mHz_Linewidth_Haonan_2020}. Implementing dark atom--photon dressed states~\cite{Ultranarrow_Superradiant_Lasing_Zhang_Yuan,sundarSqueezingMultilevelAtoms2024} on the $^{88}$Sr $^1\!S_0\to{}^3\!P_1$ transition could suppress the dominant dephasing channels and push the emission linewidth below the two-level Purcell limit, opening a route towards fractional frequency instabilities at the $10^{-16}/\sqrt{\tau/\mathrm{s}}$ level. Extending continuous SR lasing to true clock transitions, namely the ultra-narrow $^1\!S_0\to{}^3\!P_0$ lines in alkaline-earth(-like) atoms, could provide short-term frequency stability comparable to that of the best optical reference cavities~\cite{elbadawiFrameworkContinuousSuperradiant2026,zhangActiveFrequencyMeasurement2022,reillyFullyCollectiveSuperradiant2026a,dubeyModelingContinuousSuperradiant2025,Continuous_Collective_Cline_2025}, while simultaneously serving as an atomic frequency reference.
As SR lasers are less sensitive to mechanical and thermal noise than cavity-stabilized lasers, they are natural candidates for transportable optical clocks~\cite{Mehlstaubler2018}, either as \emph{active} optical clocks in their own right or in combination with a conventional \emph{passive} optical clock, as is done with the hydrogen maser and caesium fountain clock in the microwave domain~\cite{Parker_hydrogen_masers_2010}. 
Both strategies offer a promising path towards the next generation of optical clocks for precision metrology.
Beyond metrology, we have begun to explore complex collective dynamics, observing coherent collective amplitude oscillations, a characteristic signature of multi-component superradiance. These dynamics open promising avenues for quantum simulation of driven-dissipative phase transitions, including non-Hermitian phase transitions, spontaneous symmetry breaking, and the emergence of chaos~\cite{yi-xiangGoldstoneHiggsModes2013,mivehvarDrivenDissipativeSupersolidRing2018,kirtonIntroductionDickeModel2019,ozturkObservationNonHermitianPhase2021}.

\clearpage
\section*{Methods}

\subsection*{Experimental apparatus and calibration}
\label{sec:methode_1}

\begin{figure*}[t]
\extendeddatafigures
\centering
\includegraphics[width=\textwidth]{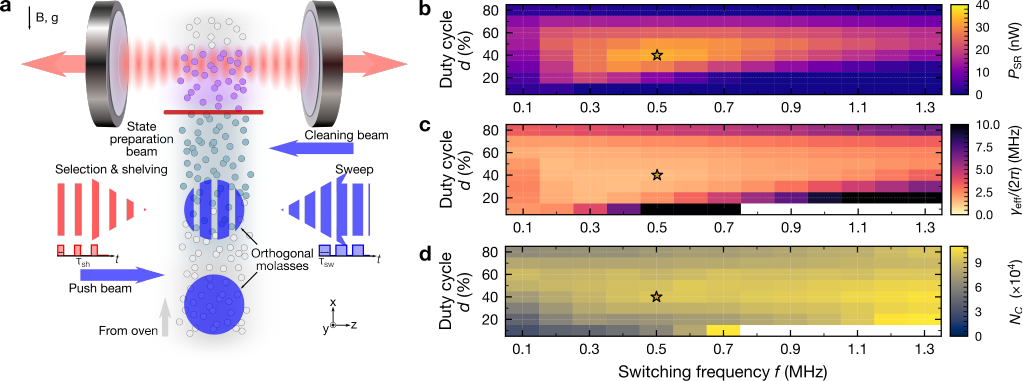}
\caption{
\textbf{Apparatus and parameter space of continuous superradiant emission.}
\textbf{(a)}~Schematic of the experimental apparatus showing the atomic beam propagating vertically against gravity, the SVS cooling zone, the cleaning zone, the state preparation beam, and the optical cavity (SR cavity). The directions of the magnetic field $B$ and gravity $g$ are indicated.
\textbf{(b)}~Mean steady-state SR output power $P_\mathrm{SR}$ as a function of switching frequency $f$ and duty cycle $d$.
\textbf{(c)}~Effective linewidth $\gamma_\mathrm{D}\sim\gamma_\mathrm{eff}$ extracted from NMS fits [Eq.~\eqref{eq:gammaD}].
\textbf{(d)}~Intra-cavity atom number $N_C = (G/g_0)^2$ [Eq.~\eqref{eq:NC_cavity}] extracted from the same NMS fits, with $g_0/2\pi = 11.3\,\mathrm{kHz}$. The reference operating point is indicated with a star.
For (b)-(d) each data pixel is the mean of four independent measurements.}
\label{fig:Figure_1_methods}
\end{figure*}

\textbf{Atomic source and cavity setup.}
A full description of the apparatus, the cooling scheme, and its characterization by normal-mode-splitting (NMS) measurements is given in Ref.~\citenum{famaCollimationDenseAtomic2026}.
A schematic of the apparatus is shown in Ext. Data Fig.~\ref{fig:Figure_1_methods}(a). A high-flux oven produces a thermal beam of ${}^{88}$Sr atoms. The SR cavity is mounted $\sim15\,\mathrm{cm}$ above the oven nozzle, with its axis oriented orthogonal to the atomic beam. The cavity has length $L = 26.65$\,mm, finesse $\mathcal{F} = 828(2)$ and linewidth (FWHM) $\kappa/2\pi = 6.79(2)\,\mathrm{MHz}$, measured at the atomic transition wavelength $\lambda = 689\,\mathrm{nm}$. The waist of the cavity mode, $w_0 = 87\,\mathrm{\mu m}$, is estimated from the cavity design and yields a maximum single-atom coupling rate
\begin{equation}
    g_0/{2\pi}= -\frac{\hat{\varepsilon}\cdot\mathrm{d}}{h}\sqrt{\frac{\omega\hbar}{2\epsilon_0 V}}
    = 11.3\,\mathrm{kHz},
    \label{eq:g0}
\end{equation}
where $\hat{\varepsilon}\cdot\mathrm{d}$ is the ${}^1\!S_0 \leftrightarrow {}^3\!P_1,\,m_J=0$ dipole matrix element projection, $\omega = 2\pi c/\lambda$ is the transition frequency and $V=\pi w_0^2 L/4$ is the effective mode volume, accounting for the Gaussian transverse intensity profile and the standing-wave mode function~\cite{Single-atom_detection_Poldy_2008}. The single-atom coupling is used to extract the atom number from NMS measurements and to define the superradiant lasing threshold. The oven nozzle consists of a rectangular array of collimation tubes (12\,mm $\times$ 2\,mm), elongated along the cavity direction. The oven is operated at $525\,^{\circ}\mathrm{C}$, which, in the Knudsen diffusion regime, corresponds to a mean atomic transverse velocity of $\bar{v}_x \approx 345\,\mathrm{m/s}$. The TEM$_{00}$ cavity mode has a Gaussian transverse field profile $\mathcal{E} \propto \exp(-x^2/w_0^2)$, so that the effective mean interaction time, defined by the pulse-area integral $\Theta = \Omega_0\tau$, is
\begin{equation}
    \tau_T = \frac{\sqrt{\pi}\,w_0}{\bar{v}_x} \approx 0.45\,\mu\mathrm{s},
    \label{eq:tau}
\end{equation}
a factor $\sqrt{\pi}/2 \approx 0.89$ shorter than the flat-top estimate $\tau_\mathrm{flat} = 2w_0/\bar{v}_x \approx 0.50\,\mu\mathrm{s}$, which provides an upper bound on the interaction time. The resulting transit-time broadening is
\begin{equation}
    \gamma_{\mathrm{T}} = \frac{1}{\tau_T}
    = \frac{\bar{v}_x}{\sqrt{\pi}\,w_0}
    \approx 2\pi \times 360\,\mathrm{kHz}.
    \label{eq:gamma_T}
\end{equation}
%
%
The collimation from the nozzle array limits the transverse velocity spread to $|v_z| \lesssim 10\,\mathrm{m/s}$ if we assume no wall collisions in the tubes. This corresponds to a Doppler broadening of $\gamma_\mathrm{D} = 2\pi \times v_z/\lambda \lesssim 2\pi \times 14.5\,\mathrm{MHz}$, far exceeding that allowed by the superradiant lasing threshold condition.

\textbf{Laser cooling and state preparation.}
\label{sec:SVS}
Reducing $\gamma_\mathrm{D}$ to below the superradiant lasing threshold requires laser cooling parallel to the cavity axis. 
We employ Swept Velocity Shelving (SVS), explained in detail in Ref.~\citenum{famaCollimationDenseAtomic2026}.
The SVS selection beam is Gaussian (waists $w_x = 35\,\mathrm{mm}$, $w_y = 0.8\,\mathrm{mm}$) and power-broadened at its peak to a linewidth of $\approx 900\,\mathrm{kHz}$. Compared to the top-hat beam that was used for Ref.~\citenum{famaCollimationDenseAtomic2026}, the Gaussian beam gives slightly higher transfer efficiency at comparable Doppler broadening.
Above the SVS zone, a tightly focused \textit{cleaning beam}, propagating in the same direction as the sweep beam and resonant with ${}^1\!S_0 \rightarrow {}^1\!P_1$, Doppler shifts residual ground-state atoms out of resonance, suppressing their absorption of intra-cavity photons.
In the final \textit{state preparation} stage, atoms shelved in ${}^3\!P_{0,2}$ during SVS are repumped via ${}^3\!S_1$ and decay spontaneously into the ${}^3\!P_1$ manifold. The repump light is delivered in a sheet beam propagating orthogonal to both the atomic beam and the selection beam, with a waist $w_x= w_\mathrm{SP} =  150\,\mu\mathrm{m}$ chosen to place the excitation region close to the cavity mode while avoiding light shifts inside it. 
To maximize the population in the target state ${}^3\!P_1,\,m_J = 0$, we include additional frequency components
addressing ${}^3\!P_1,\,m_J = \pm 1 \rightarrow {}^3\!S_1$, giving atoms in ${}^3\!P_1,\,m_J = \pm 1$ a further chance to decay into ${}^3\!P_1,\,m_J = 0$ and contribute to the superradiant emission.

\textbf{Atom-cavity interaction and normal-mode splitting.}
To characterize the intra-cavity atom number and Doppler broadening, we use normal-mode splitting (NMS)~\cite{Thompson1992ObservationNormalMode,Raizen1989NormalModeSplitting,Zhu1990VacuumRabiSplitting} by ground-state atoms. For this measurement, the state preparation beam is positioned $\Delta x = 1.5\,\mathrm{cm}$ below the SR cavity mode, giving atoms sufficient time to decay spontaneously from ${}^3\!P_1$ to ${}^1\!S_0$ before reaching the SR cavity. At our mean beam velocity $\bar{v}_x = 345\,\mathrm{m/s}$, more than $80\%$ of atoms have decayed to the ground state by the time they enter the SR cavity mode. The ground-state atoms then couple collectively to the cavity field, and the collective coupling strength
\begin{equation}
    G = g_0\sqrt{N_C}
    \label{eq:G}
\end{equation}
becomes comparable to or exceeds the SR cavity decay rate $\kappa$, causing the cavity transmission to split into two hybridized normal modes. The intra-cavity atom number $N_C$ and the effective atomic linewidth $\gamma_\mathrm{eff}$ are extracted simultaneously by fitting the measured transmission spectrum with the general atom-cavity transmission model, Ref.~\citenum{famaCollimationDenseAtomic2026}, where the cavity linewidth $\kappa$ is fixed from independent bare-cavity measurements in the absence of atoms.
We assume the ground- and excited-state populations differ by only one spontaneously emitted photon, so that the NMS measurement on ground-state atoms provides a faithful map of the velocity distribution and atom number of the atoms that will subsequently contribute to superradiant emission. From the cavity transmission fit, the effectively coupled intra-cavity atom number is obtained as
\begin{equation}
    N_C = \left(\frac{G}{g_0}\right)^2.
    \label{eq:NC_cavity}
\end{equation}
Homogeneous broadening is characterized by the spontaneous emission at rate $\gamma$, stray-light-induced dephasing at rate $\gamma_B$, and finite interaction times with the cavity mode at rate $\gamma_\mathrm{T}$. The total effective broadening is therefore
\begin{equation}
    \gamma_\mathrm{homo} = \gamma + \gamma_B + \gamma_\mathrm{T}
    \approx 2\pi \times 400\,\mathrm{kHz},
    \label{eq:gamma_homo}
\end{equation}
larger than the value quoted in Ref.~\citenum{famaCollimationDenseAtomic2026}, where the flat-top transit-time estimate $\tau_\mathrm{flat} = 2w_0/\bar{v}_x$ was used. The Doppler broadening is isolated by subtracting the homogeneous contributions,
\begin{equation}
    \gamma_\mathrm{D} = \gamma_\mathrm{eff} - \gamma_\mathrm{homo}.
    \label{eq:gammaD}
\end{equation}
Because the SVS scheme selects atoms within a Lorentzian velocity class, both the homogeneous and Doppler contributions are well approximated by Lorentzian profiles, and their linewidths add directly~\cite{famaCollimationDenseAtomic2026}.

\textbf{Main tuning parameters.}
The superradiant emission is controlled via four main experimental parameters. The switching frequency $f = 1/(\tau_\mathrm{sel} +
\tau_\mathrm{sw})$ and duty cycle $d = \tau_\mathrm{sel} \cdot f$ set the relative duration of the selection and sweep pulses in the SVS scheme, and together determine the maximum $N_C$ and $\gamma_\mathrm{D}$. The power of the state preparation beam controls the fraction of atoms repumped into ${}^3\!P_1,\,m_J = 0$, and its position $\Delta x$ relative to the SR cavity mode sets the excited-state fraction $f_e$ at the SR cavity entrance via the natural decay of ${}^3\!P_1$ ($\gamma/2\pi = 7.48\,\mathrm{kHz}$). Note that changing the state preparation beam power does not change $f_e$, since atoms that are not repumped remain shelved in the metastable ${}^3\!P_{0,2}$ states.

\textbf{Atom number $N_C$ and Doppler broadening $\gamma_\mathrm{D}$.}
To map out the accessible parameter space, we perform NMS measurements over a two-dimensional grid of $f$ and $d$, recording
$\gamma_\mathrm{D}$ and $N_C$ at each point (Ext. Data Fig.~\ref{fig:Figure_1_methods}c-d), alongside the steady-state SR output power $P_\mathrm{SR}$ emitted through one of the cavity mirrors, i.e. half the total output power
(Ext. Data Fig.~\ref{fig:Figure_1_methods}b). The steady-state output power is measured for a duration of $100\,\mathrm{ms}$, $100\,\mathrm{ms}$ after switch-on to ensure the system has reached steady state, using a free-space silicon avalanche photodetector (APD) (Thorlabs APD130A/M).
The SR output is non-zero over a broad region of the $(f, d)$ plane, demonstrating that continuous superradiant emission is robust to variations in the cooling parameters. Within this region, the output power reaches a maximum, measured on one side of the cavity, of $32.8(1)\,\mathrm{nW}$ near $f = 0.5\,\mathrm{MHz}$, $d = 40\%$, the reference operating point that we always use unless stated otherwise. This maximum coincides with the minimum Doppler broadening $\gamma_\mathrm{D} \approx 2\pi \times 1\,\mathrm{MHz}$ found from NMS at similar parameters ($f = 0.4\,\mathrm{MHz}$, $d = 40\%$). This agreement indicates that the NMS diagnostic on ground-state atoms faithfully reflects the velocity distribution of the excited-state emitters.
The intra-cavity atom number remains approximately constant at $N_C \approx 0.98 (5) \times 10^5$ over the range $f \in [0.4, 1.3]\,\mathrm{MHz}$ at $d = 40\%$ (Ext. Data Fig.~\ref{fig:Figure_1_methods}d), providing a stable window in which $\gamma_\mathrm{D}$ can be varied, nearly independently of $N_C$. This window includes our reference operating point ($f = 0.5\,\mathrm{MHz}$). The corresponding total atomic flux through the SR cavity mode is $\Phi = 2N_C \bar{v}_x / (\sqrt{\pi}w_0) \approx 4.4(1)\times 10^{11}\,\mathrm{s^{-1}}$, reduced compared to Ref.~\citenum{famaCollimationDenseAtomic2026} since now most atoms decay through ${}^3\!P_1,\,m_J = 0$.

\begin{figure*}[t]
\centering
\includegraphics[width=\textwidth]{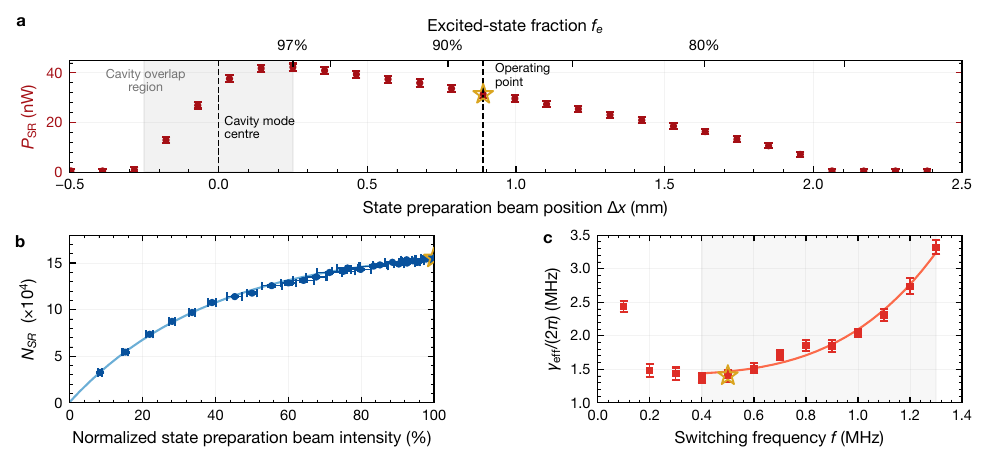}
\caption{
\textbf{Calibration of the state preparation beam position, the excited-state atom number, and the Doppler broadening.}
\textbf{(a)}~Steady-state SR output power $P_\mathrm{SR}$ as a function of state preparation beam position $\Delta x$ relative to the estimated position of the SR cavity mode centre. The corresponding excited-state fraction $f_e(\Delta x)$ is shown on the top axis [Eq.~\eqref{eq:fe}]. The grey shaded region ($|\Delta x| < w_0 + w_\mathrm{SP} \approx 0.25\,\mathrm{mm}$) marks positions where the beam overlaps the intra-cavity field.
\textbf{(b)}~Effective excited-state emitter number $N_\mathrm{SR} = 2\nu N_C$ as a function of normalized state preparation beam intensity [Eq.~\eqref{eq:nc_sat}] (solid curve).
\textbf{(c)}~Effective linewidth $\gamma_\mathrm{eff}\sim\gamma_\mathrm{D}$ as a function of switching frequency $f$ at fixed duty cycle $d = 40\%$ and maximum state preparation beam intensity. The solid curve is the empirical power-law calibration [Eq.~\eqref{eq:gd_power}], with the fit range $f \in [0.4, 1.3]\,\mathrm{MHz}$ indicated by the shaded region. Values taken at the reference operating point are marked by stars.
Error bars in (a) are defined as in Fig.~\ref{fig:Figure_1}d.
Error bars in (b) and (c) indicate the sample standard deviation of fitted values across repeated measurements at each scan point, combined in quadrature with the propagated fit uncertainty of the mean.
}
\label{fig:Figure_2_methods}
\end{figure*}

\textbf{Excitation beam position and excited-state fraction.}
To adjust the excitation fraction $f_e(\Delta x)$, we position the state preparation beam at a distance $\Delta x$ upstream of the cavity mode. Atoms then decay spontaneously from ${}^3\!P_1$ during transit to the SR cavity, giving an excited-state fraction
\begin{equation}
    f_e(\Delta x) = \exp\!\left(
    -\frac{\gamma\,\Delta x}{\bar{v}_x}
    \right).
    \label{eq:fe}
\end{equation}
We scan the position of the excitation beam $x_\mathrm{ex}$ along the atomic beam axis using a DC servo-motorized mirror mount (Thorlabs Z912), see Ext. Data Fig.~\ref{fig:Figure_2_methods}a. The SR output power peaks at $x_\mathrm{peak}$, determined with a precision of $\approx 0.05$\,mm. We believe that at this position the excitation beam is located just below the SR cavity mode, maximizing the number of excited state atoms that have the full transit time to emit a photon. We therefore estimate that the centre of the cavity mode lies $\approx w_0 + w_\mathrm{SP} \approx 0.25\,\mathrm{mm} \equiv \delta x_\mathrm{centre}$ above the peak. Using the estimated location of the cavity centre as the origin, we obtain $\Delta x(x_\mathrm{ex}) = x_\mathrm{peak} - x_\mathrm{ex} + \delta x_\mathrm{centre}$.
The reference operating point, $\Delta x^\mathrm{op} = 0.89(5)\,\mathrm{mm}$, is chosen sufficiently far from the SR cavity to avoid measurable light shifts while maintaining $f_e^\mathrm{op} \simeq 89\%$.
The effective emitter number is approximately 
\begin{equation}
    \nu = \eta_\mathrm{repump} \cdot f_e^\mathrm{op}
    \approx 0.8,
    \label{eq:nu}
\end{equation}
where $\eta_\mathrm{repump} \approx 0.90$ is the repumping efficiency into ${}^3\!P_1,\,m_J = 0$. The dominant uncertainty arises from $\eta_\mathrm{repump}$. As it enters only as an overall linear scaling of $N_\mathrm{SR}$ it does not affect the threshold behaviour or the shape of the transition.

\textbf{Effective emitter number $N_\mathrm{SR}$.}
The effective emitter number $N_\mathrm{SR}$ is the total number of excited-state atoms inside the $1/e$ cavity mode volume, i.e. without standing-wave weighting. It is calculated from the NMS atom number $N_C$, which includes the axial average, $\langle \cos^2(kz) \rangle = 1/2$, over the standing-wave mode function. We calculate the effective emitter number as
\begin{equation}N_\mathrm{SR} = 2\nu N_C,
    \label{eq:Nsr}
\end{equation}
where $\nu \approx 0.8$ accounts for the excited-state fraction and repumping efficiency [Eq.~\eqref{eq:nu}]. 

We measure $N_C$ via NMS as a function of the excitation beam intensity under the reference operating conditions $f = 0.5\,\mathrm{MHz}$ and $d = 40\%$, see Ext. Data Fig.~\ref{fig:Figure_2_methods}b. The resulting $N_\mathrm{SR}(I)$ dependence is well described by
\begin{equation}
    N_\mathrm{SR}(I) = 2\nu\left(a_N\!\left(1 - e^{-I/b_N}\right) + c_N\right),
    \label{eq:nc_sat}
\end{equation}
with fit results $a_N = 1.040(5) \times 10^5$, $b_N = 39.0(8)\%$, $c_N = 670(530)$ and the fit quality characterized by a reduced $\chi^2 = 0.9$. $I$ is the peak preparation beam intensity normalized to 125\,mW/cm$^2$~\cite{famaCollimationDenseAtomic2026}.

\textbf{Doppler broadening calibration.}
At fixed duty cycle $d = 40\%$ and maximum intensity in the state preparation beam, the switching frequency $f$ is the last control parameter. As established above, $N_C$ remains approximately constant at $\langle N_C \rangle \approx 0.98 (5) \times 10^5$ over the range $f \in [0.4, 1.3]\,\mathrm{MHz}$, so that $N_\mathrm{SR} \approx 2\nu \langle N_C \rangle \approx 1.57(8) \times 10^5$ atoms is treated as constant. The Doppler broadening $\gamma_\mathrm{D}=\gamma_\mathrm{eff}-\gamma_\mathrm{homo}$, where $\gamma_\mathrm{eff}$ is measured with NMS and $\gamma_\mathrm{homo}$ is assumed to be constant, varies strongly with $f$ and is described by the empirical power law
\begin{equation}
    \gamma_\mathrm{D}(f) = a_\gamma\,(f/\mathrm{MHz})^{b_\gamma} + c_\gamma,
    \label{eq:gd_power}
\end{equation}
as shown in Ext. Data Fig.~\ref{fig:Figure_2_methods}c. The fit results in $a_\gamma= 0.67(8)$\,MHz, $b_\gamma = 3.8(4)$ and $c_\gamma = 1.42(6)$\,MHz with a reduced $\chi^2 = 1.38$. The fitted exponent $b_\gamma$ reflects the nonlinear dependence of the selected velocity class on the sweep/selection pulse duration in the SVS scheme~\cite{famaCollimationDenseAtomic2026}. For all experiments described in the main text, except the MCSR measurements, the duty cycle is fixed at $d = 40\%$, the excitation beam at $\Delta x^\mathrm{op} = 0.890\,\mathrm{mm}$, and we vary the state preparation beam intensity to tune $N_\mathrm{SR}$ and vary $f$ to tune $\gamma_\mathrm{D}$.

\subsection*{Optical setup and frequency stability}
The frequency stability and accuracy characterization is based on three lasers: the superradiant laser itself (SR laser), a ULE-cavity-stabilized reference laser (ULE laser), and a spectroscopy laser (Spec laser). The spectroscopy laser is stabilized on short timescales ($<1\,\mathrm{s}$) to a transfer cavity, limiting its linewidth to a few kHz, and on longer timescales to the ${}^{88}$Sr ${}^1\!S_0 \rightarrow {}^3\!P_1$ transition via modulation-transfer spectroscopy on a hot vapour cell~\cite{StellmerPhD2013,qiao_ultrastable_2019}. Because its linewidth far exceeds that of the ULE laser, the spectroscopy laser serves only as a long-term frequency reference.
The ULE laser is stabilized to a ULE cavity (Stable Lasers Ltd.~\cite{leibrandtSphericalReferenceCavities2011}) with a zero-crossing temperature of approximately $5\,^\circ\mathrm{C}$. When operated near this temperature, the cavity drift rate is $290 (10)\,\mathrm{Hz/h}$, sufficiently low that the beat linewidth is limited by fibre-link noise rather than cavity drift. This condition is used for the intrinsic linewidth characterization of the SR laser. To demonstrate the self-locking capability of the SR laser, we deliberately detune from the zero-crossing temperature, increasing the drift rate to $4215 (1)\,\mathrm{Hz/h}$. Under these conditions the SR self-lock is engaged to suppress the cavity drift of the interaction cavity (Fig.~\ref{fig:Figure_3}a).

\begin{figure*}[t]
\centering
\includegraphics[width=\textwidth]{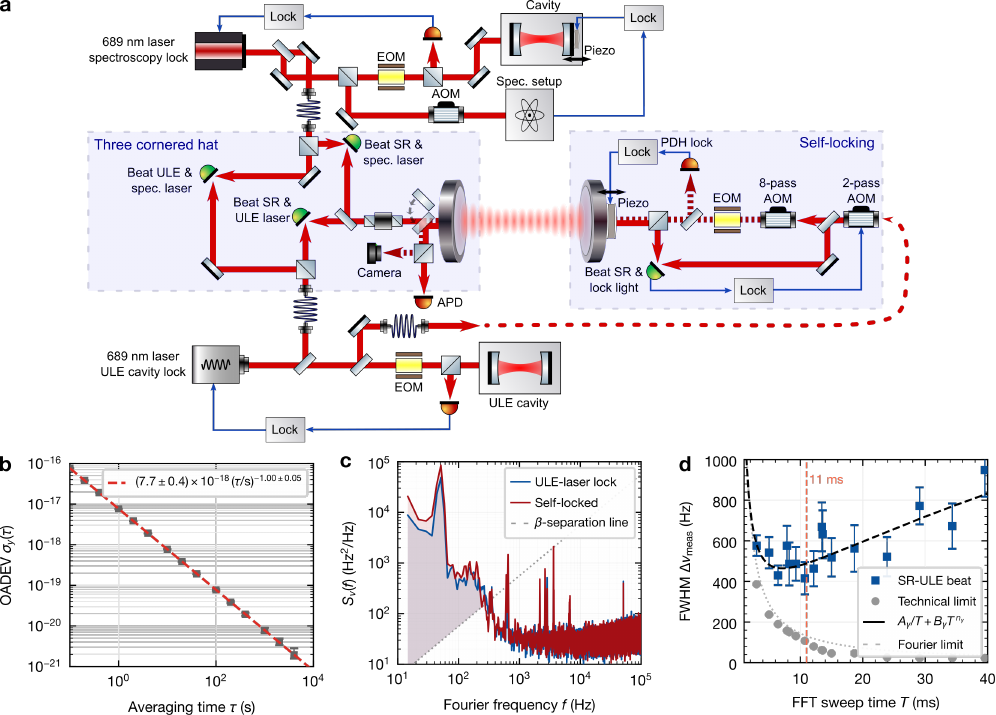}
\caption{
\textbf{Linewidth characterization and self-locking of the superradiant laser.}
\textbf{(a)}~Schematic of the optical setup. The SR laser emits through both cavity mirrors. A flip mirror on the left selects between amplitude measurement and three-cornered hat heterodyne frequency measurement. The right cavity output is used for the self-locking scheme, forcing the cavity length to follow the SR laser frequency.
\textbf{(b)} Overlapping Allan deviation of the SR - lock light beat note, while all locks are engaged (in-loop), with power law fit (red line).
\textbf{(c)}~Frequency noise power spectral density $S_\nu(f)$, over 1 s measurement period, for which the SR cavity is locked directly to the ULE laser (blue) or is self-locked (red). The dotted grey line shows the $\beta$-separation line $\beta f$ ($\beta = 8\ln 2/\pi^2 \approx 0.56$). Here, the shaded areas indicate the noise contributions to the $\beta$-separation linewidth~\cite{DiDomenico2010}. No drift subtraction is applied and uncertainties are estimated by bootstrapping ($n = 200$ Welch-segment resamples).
\textbf{(d)}~Intrinsic linewidth $\Delta\nu_\mathrm{meas}$ versus FFT sweep time $T$, measured via a heterodyne beat with the ULE laser (blue squares). Grey disks show the technical noise floor extracted from an RF reference signal whereas the grey dotted line is the Fourier limit $1.44/T$. The black dashed curve is a fit to [Eq.~\eqref{eq:FWHMfit}]. The sweep time for all laser linewidth measurements mentioned elsewhere is marked by the orange dashed line.
Error bars: (b), as in Fig.~\ref{fig:Figure_3}b; (d), as in Fig.~\ref{fig:figure2}.}
\label{fig:Figure_3_methods}
\end{figure*}

\textbf{Optical setup and heterodyne detection.}
The optical setup is schematically shown in Ext. Data Fig.~\ref{fig:Figure_3_methods}a. The SR cavity has equal mirror coatings on both sides, resulting in bidirectional emission of equal power. The left output is directed to the detection breadboard.
On the detection breadboard, a flip mirror selects between two measurement configurations: amplitude detection on a free-space APD (bandwidth 50\,MHz) where the SR amplitude is measured directly, or frequency measurement using a heterodyne three-cornered hat setup. This separation improves the signal-to-noise ratio for both measurements.
In the three-cornered hat configuration, the SR output is first sent through a free-space optical isolator, which prevents back-scattered light from the ULE laser or Spec laser from coupling into the SR cavity and disturbing emission, and then split equally and directed to two heterodyne beat measurements. The first beat is formed between the SR laser and the ULE laser (ULE-SR, $f_\mathrm{beat} = 3.3$\,MHz), the second between the SR laser and the Spec laser (SR-Spec, $f_\mathrm{beat} = 10.9$\,MHz) and detected via free-space silicon avalanche photodetectors. A third beat between the ULE laser and the Spec laser (ULE-Spec, $f_\mathrm{beat} = 7.6$\,MHz) is detected on a free-space silicon photodiode on the same breadboard. All three beats are detected with a bandwidth of 50\,MHz and isolated with RF bandpass filters of 3--5\,MHz bandwidth centred on the respective beat frequencies. 

Because the Spec laser's linewidth far exceeds that of the ULE laser, the former is excluded from short-term linewidth measurements; short-term characterization relies on the ULE-SR beat alone. The quoted linewidths are upper bounds on the intrinsic SR linewidth. For long-term stability  characterization, all three beats are recorded simultaneously with a K\&K FXE phase and frequency counter, allowing a three-cornered hat comparison that isolates the individual frequency noise contribution of each laser~\cite{vernotteThreecorneredHatAllan2016}. For the strongly drifting ULE laser, shown in Fig.~\ref{fig:Figure_3}b, we applied a linear drift removal before calculating the OADEV. 

We additionally analyse the long-term stability shown in Fig.~\ref{fig:Figure_3}. Over the measurement duration, the mean frequency offset between the spectroscopy laser and the self-referenced SR laser is $77.5(2.3)$\,kHz (SR-Spec), where the uncertainty is the standard deviation of the offset over the full measurement. The offset may originate from the spectroscopy setup, from imperfect centring of the cavity on the transition frequency, or from shifts induced by the lock light and the intracavity atom number, as discussed in SI.

\textbf{Cavity length stabilization and self-locking scheme.}
The length of the SR cavity is stabilized onto the SR laser frequency itself, which is possible because of the small cavity-pulling coefficient. As the SR laser output does not provide enough power to implement the cavity lock, we first beat-lock a portion of the ULE laser light to a fixed offset frequency from the SR laser (3\,MHz), see "Self-locking" breadboard in Ext. Data Fig.~\ref{fig:Figure_3_methods}a. Any slow drift of the ULE cavity is thereby absorbed by the acousto-optical modulator (AOM) feedback loop (bandwidth 100\,kHz), and the cavity length stabilization laser beam follows the SR laser frequency rather than the ULE cavity, as confirmed by the $\tau^{-1}$ scaling of the in-loop Allan deviation of the SR - lock light beat note (Ext. Data Fig.~\ref{fig:Figure_3_methods}b). Furthermore, this SR cavity lock setup keeps the approximation of three independent oscillators in the three cornered hat valid, despite using light from the ULE laser for the SR cavity lock (100\,kHz bandwidth of the lock vs 100\,ms $\tau$ gate time in the measurement). We use self-locking for the long-duration measurements shown in Fig.~\ref{fig:Figure_3}a and Ext. Data Fig.~\ref{fig:Figure_3_methods}b. For short-duration measurements the self-locking scheme is not beneficial and, as stated for each such measurement, we lock directly to ULE laser light shifted by a fixed frequency. 

The cavity lock light is red-shifted to $-2650$\,MHz of the ${}^1\!S_0 \rightarrow {}^3\!P_1$ transition via an 8-pass AOM~\cite{Zhou_Laser_Frequency_Shift_2020}, placing it on resonance with a TEM$_{02}$ cavity mode. The TEM$_{02}$ mode is chosen to separate the frequencies of the locking light and the SR emission as well as for its reduced spatial overlap with the lasing TEM$_{00}$ mode. Sideband modulation at 50\,MHz is applied via a fibre electro-optical modulator (EOM), and a standard Pound-Drever-Hall (PDH) scheme generates the cavity length error signal. To reduce the influence of the locking light as much as possible, a free-space silicon avalanche photodetector was utilized as a PDH photodiode, allowing the locking light power to be reduced to well below 1\,$\mu$W. The linear locking light polarization is set orthogonal to the linear SR emission polarization allowing the lock light to be separated from the SR output on the detection side using a polarizing beam splitter (PBS). The cavity transmission is monitored with a CCD camera.

The cavity linewidth is practically the same for the lasing and the locking mode ($\kappa/2\pi \approx 6.79$\,MHz), resulting in a PDH error signal-to-noise ratio (SNR) of only 500-1000. This limits the cavity lock precision to $\sim$tens of kHz. However, the influence of this cavity frequency uncertainty on the SR laser frequency is suppressed by the small bad-cavity pulling coefficient $\rho = 0.165 (3)$, reducing its contribution to the SR frequency uncertainty to $\sim$few kHz. We chose a slow cavity piezo lock bandwidth of $\sim$10\,Hz as the cavity exhibits excellent passive thermal and vibrational stability~\cite{BeliSilva2025CavityAtom,Fama:25}. Increasing the lock bandwidth would degrade, rather than improve, the SR frequency stability on the 10\,ms timescale relevant for linewidth extraction.

After self-locking, an external frequency reference remains in the system: the SVS selection laser, which is derived from the Spec laser. A drift of its frequency displaces the velocity class selected along the cavity axis, but does not pull the SR emission frequency. Because each atom couples to both counterpropagating components of the cavity field, the effective velocity distribution along the cavity axis is symmetrized. Consequently, a nonzero mean velocity can appear as a symmetric splitting of the gain profile rather than as a shift. This splitting is resolved only when it exceeds the gain broadening, making this signature robust against velocity offsets~\cite{famaContinuousCavityQED2024}. The emission frequency is therefore first-order insensitive to drifts of the selection laser, which also ensures the independence of the SR and Spec lasers assumed in the three-cornered-hat analysis. In the future, the selection laser could be derived from the SR laser, removing the need for the spectroscopy cell.

\textbf{Short-term linewidth.}
Two complementary quantities are used to characterize the SR laser frequency noise, each giving access to a different aspect of the emission. The frequency noise power spectral density $S_\nu(f)$ captures the full noise spectrum including lock-induced contributions and characterizes the laser as seen by an external user on a one second timescale by the $\beta$-separation linewidth $\Delta\nu_\beta$ (see Ext. Data Fig.~\ref{fig:Figure_3_methods}c and Supplementary Information). The short-term intrinsic linewidth $\Delta\nu_\mathrm{meas}$ describes short-term phase stability. 
This linewidth is extracted at timescales shorter than the cavity lock bandwidth ($\sim$10\,Hz), where the lock has not yet acted and the SR emission is effectively free-running.
To access the intrinsic linewidth $\Delta\nu_\mathrm{meas}$, we follow the method of Ref.~\citenum{Kristensen2023} and restrict the measurement to FFT sweep times $T \ll 1/f_\mathrm{lock} = 100$\,ms.
Here, we use the ULE laser directly to stabilize the SR cavity length because of the slightly lower $\Gamma_{\beta_\mathrm{ULE}}$. The ULE-SR beat signal is recorded via the oscilloscope in FFT mode (Hann window) and a Lorentzian is fitted within a 10-kHz window centred on the beat frequency for each sweep time $T$:
\begin{equation}
    L(f) = A \frac{(\Delta\nu_\mathrm{meas}/2)^2}{(f-f_c)^2 + 
    (\Delta\nu_\mathrm{meas}/2)^2} + B,
\end{equation}
where $\Delta\nu_\mathrm{meas}$ is the FWHM linewidth and $B$ is the noise floor. Each data point represents the mean of 20 independent measurements. The measured linewidth follows 
\begin{equation}
    \Delta\nu_\mathrm{meas}(T) = A/T + BT^n,
    \label{eq:FWHMfit}
\end{equation}
with fit parameters  $A = 1220 (250)$\,Hz$\cdot$ms, $B = 93 (40)$\,Hz/ms$^n$, $n = 0.58 (13)$, having a minimum at time $T_\mathrm{min} = 7.1$\,ms (Ext. Data Fig.~\ref{fig:Figure_3_methods}d). For all laser
linewidth measurements mentioned elsewhere, we choose an FFT sweep time of $T_\mathrm{FFT} = 11.0$\,ms. This conservative choice provides reproducible measurements and simultaneously minimizes the Fourier limit ($1.44/T_\mathrm{FFT} = 130$\,Hz) and the technical noise floor ($107.5 (6)$\,Hz from the RF reference). At $T_\mathrm{FFT}$ this data set yields its minimum measured linewidth of $\Delta\nu_\mathrm{meas} = 420 (80)$\,Hz. The increase at longer sweep times reflects technical noise broadening. 

\begin{figure}[t]
\centering
\includegraphics[width=\singlecolwidth]{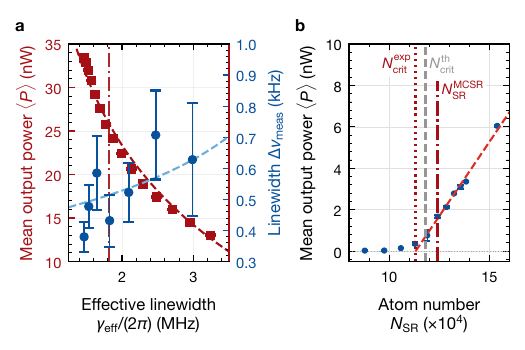}
    \caption{
    \textbf{Mean SR output power $\langle P \rangle$ and intrinsic linewidth during transition from single frequency to multi-component lasing.}
    \textbf{(a)} Mean output power $\langle P \rangle$ (red squares, left axis) and intrinsic SR laser linewidth $\Delta\nu_\mathrm{meas}$ of the carrier (blue disks, right axis) as a function of $\gamma_\mathrm{eff}$. $\Delta\nu_\mathrm{meas}$ is extracted from a multi-peak Lorentzian fit to the spectrum and represents an upper bound on the intrinsic SR laser linewidth. The analytical model Eq.~\eqref{eq:linewidth_main} is fitted to the linewidth measurements, while adding a constant offset to account for the finite probe laser linewidth and technical noise contributions ($\Delta\nu_\mathrm{add} = 400(40)$\,Hz), shown as the dashed blue line. The output power model Eq.~\eqref{eq:Power_out} (red dashed line) yields $N_\mathrm{SR} = 1.6(1)\times10^5$ atoms. The transition to multi-component lasing happens at $\gamma_\mathrm{eff}^\mathrm{MCSR} = 1.8(1)$\,MHz (vertical red dash-dotted line).
    \textbf{(b)} Mean SR output power $\langle P \rangle$ for a large broadening $\gamma_\mathrm{eff}/(2\pi) = 4.5 (1)$\,MHz as a function of $N_{\rm SR}$. A linear fit above threshold (red dashed line) yields $N_\mathrm{crit}^\mathrm{exp} = 1.13(1)\times10^5$ (vertical red dotted line). The theoretical threshold $N_\mathrm{crit}^\mathrm{th} = 1.18\times10^5$ is indicated with a vertical grey dashed line. The transition to multi-component lasing happens at $N_\mathrm{SR}^\mathrm{MCSR} = 1.24(5)\times10^5$ (vertical red dash-dotted line). Error bars: (a), as in~Fig.~\ref{fig:figure2}b; (b), as in~ Fig.~\ref{fig:Figure_1}d. 
    }
    \label{fig:Figure_6_methods}
\end{figure}

As we measure the beat between the SR laser and the ULE-stabilized probe laser, the measured linewidth is $\Delta\nu_\mathrm{meas} = \Delta\nu_\mathrm{SR} + \Delta\nu_\mathrm{ULE}$. To account for the technical noise sources, the superradiant cavity lock and the finite probe linewidth, we fit a constant offset $\Delta\nu_\mathrm{add}$ with Eq.~\eqref{eq:linewidth_main} to the data (Fig.~\ref{fig:figure2}b and Ext. Data Fig.~\ref{fig:Figure_6_methods}a), obtaining $\Delta \nu _{\rm add} = 310 (60)\,\mathrm{Hz}$ from the $N_{\rm SR}$ scan and $\Delta\nu_\mathrm{add} = 400 (40)\,\mathrm{Hz}$ from the $\gamma_{\rm eff}$ scan. The two independent fits are consistent within $1.2\,\sigma$, indicating that $\Delta\nu_\mathrm{meas}$ is mostly dominated by technical noise. Including this offset, the model predicts a measured linewidth of $\sim\!400\,\mathrm{Hz}$ at $N_\mathrm{SR} = 1.57 \times 10^5$, above the lowest measured value of $320 (40)\,\mathrm{Hz}$ (obtained from an independent 30-shot measurement), as an upper bound on the intrinsic SR linewidth.

The probe laser contribution $\Delta\nu_\mathrm{ULE}$ arises from technical noise sources. The ULE cavity itself should in the ideal case only contribute approximately $\sim\!10$\,Hz due to thermal noise when operated at the zero-crossing temperature~\cite{leibrandtSphericalReferenceCavities2011}. One relevant contribution is expected from the uncompensated 10\,m long PM fibre between the ULE cavity and the SR laser. Thermomechanical $1/f$ noise from structural damping of the fibre's glass and acoustic coupling from the laboratory environment are both well-established broadening mechanisms for optical fibre paths at this length scale~\cite{duanGeneralTreatmentThermal2012, confortiAcoustical1Noises2020, zhangThermalnoiselimitedFiberFrequency2024}. Integrating over the relevant Fourier band for our 11-ms window yields an expected contribution of $\sim\!200$-$400$\,Hz. Additional contributions from residual cavity lock noise and cavity vibrations, as discussed above, cannot be excluded. 

\textbf{Cavity pulling suppression.}
To measure the cavity-pulling coefficient, $\rho=\Delta f_\mathrm{SR}/\Delta f_\mathrm{cav}$, we directly use the ULE laser light to lock the SR cavity, detune the cavity from resonance using the eight-pass AOM and measure the resulting SR frequency shift for several cavity-detuning values. The detuning points are randomized to suppress systematic shifts caused, for example, by a slowly drifting reference laser (Fig.~\ref{fig:figure2}a). The slope of the SR emission-frequency shift $\Delta f_\mathrm{SR}$ versus the cavity detuning $\Delta f_\mathrm{cav}$ yields the cavity-pulling coefficient. We use a heterodyne beat between the SR output and the ULE laser, recorded on an avalanche photodiode and analysed with a scanning spectrum analyser (beat frequency $3.15$\,MHz, span $0.1$--$8.1$\,MHz, RBW$=10$\,kHz). A Lorentzian is fitted to the central beat feature to extract the SR frequency. For this measurement, the ULE cavity is operated close to its zero-crossing temperature and calibrated against the Spec laser; its drift during an individual scan, which lasts only a few minutes, is negligible compared with the measured frequency shifts. This method provides a wide frequency span and is used to map the pulling slope across multiple Doppler widths $\gamma_\mathrm{eff}$.

\textbf{MCSR measurement.}
To vary $\gamma_\mathrm{eff}$ at fixed intracavity atom number $N_\mathrm{SR} \approx 1.57(8) \times 10^5$, we vary the switching frequency of the SVS scheme as described earlier and shown in Ext. Data Fig.~\ref{fig:Figure_2_methods}c. To scan the intracavity atom number at fixed Doppler broadening, we vary the state-preparation beam power as shown in Ext. Data Fig.~\ref{fig:Figure_2_methods}b, operating SVS at duty cycle $d = 30\%$ and switching frequency $f = 1.1$\,MHz, which sets $\gamma_\mathrm{eff}/(2\pi) = 4.5 (1)$\,MHz at atom number $N_\mathrm{SR} = 1.54(5) \times 10^5$.
In both scans, the SR laser is allowed to reach steady state for 100\,ms before data acquisition, which is well justified given that the relaxation oscillations visible in Fig.~\ref{fig:Figure_1}c decay on a timescale of ${\sim}5\,\mu$s, more than four orders of magnitude shorter than the settling time.

Two independent observables are recorded at each operating point. First, the heterodyne beat between the SR output and the ULE cavity laser is recorded on an APD and analysed with a spectrum analyser (beat frequency 3.15\,MHz, frequency range $0.1$--$8.1$\,MHz, RBW\,=\,10\,kHz, frequency spacing 13.3\,kHz), binned by a factor of four to a final resolution of 53\,kHz. This gives access to both phase and amplitude fluctuations of the SR field, and directly reveals sidebands in the optical spectrum when the MCSR regime is entered.
Second, the SR output is monitored directly on an APD without a reference laser for 10\,ms at 90\,MSa/s on an oscilloscope. The DC component yields the mean output power $\langle P \rangle$, and the AC component $\delta P(t)$ is converted to power fluctuations. A one-sided power spectral density is computed as
\begin{equation}
    S(f) = \frac{2\,|\tilde{\delta P}(f)|^2}{f_s \sum_n w_n^2},
    \label{eq:PSD}
\end{equation}
where $\tilde{\delta P}(f)$ is the discrete Fourier transform of the Hann-windowed signal, $f_s$ is the sampling rate, and $w_n$ are the window coefficients. Frequency bins are averaged to a final resolution of 50\,kHz, similar to the one of the beat-note measurement.

\ifpralook\else

\section*{Data availability}
The datasets generated and/or analysed during the current study are available from F.S. on reasonable request.

\section*{Code availability}
The custom simulation code used for the theoretical modelling is available from the corresponding author on reasonable request.

\fi 

\section*{Acknowledgements}
We thank UvA's technical workshop for electronic and mechanical assistance, in particular Daan Giesen, Edwin Baaij, Tristan van Klingeren, Ronald Hassing, Maarten Tuijl, and Johan Mozes. We also thank Ivo Knottnerus and Alex Urech for their technical support. S.B.J. gratefully acknowledges support from Corinna Kollath.  We thank Helmut Ritsch and Robert Spreeuw for providing comments on the manuscript. 

\section*{Funding}
This work was supported by the Dutch National Growth Fund (NGF), as part of the Quantum Delta NL programme. This work has received funding from the European Union’s (EU) Horizon 2020 research and innovation programme under Grant Agreement No. 820404 (iqClock project) and No. 860579 (MoSaiQC project). A.S. and S.A.S. acknowledge support from this EU programme under Grant Agreement No. 101154998 (CCSuperClock project) and Grant Agreement No. 101109698 (DarkCavities Project), respectively. S.A.S. additionally acknowledges support from the NNF Quantum Computing Programme (NQCP) and from ESA OSIP contract No. 4000145931. S.B.J. acknowledges support from the Deutsche Forschungsgemeinschaft (DFG, German Research Foundation) under Projects DFG No.~277625399 – TRR 185 OSCAR (``Open System Control of Atomic and Photonic Matter'', B4), and DFG No.~277146847 – CRC 1238 (``Control and dynamics of quantum materials'', C05) and under Germany’s Excellence Strategy–Cluster of Excellence Matter and Light for Quantum Computing (ML4Q), Grant No. EXC 2004/1 – 390534769.

\section*{Author contributions}
B.H., F.F., C.B.S., Y.C., S.A.S. and A.S. contributed to the construction of the SR laser. B.H. and A.S. recorded and analysed the data, and prepared the figures. S.B.J. developed the theoretical model and performed the simulations, with contributions from B.H. The manuscript was written by B.H., S.B.J., S.A.S., A.S. and F.S. The project was supervised by F.S. with support from S.A.S. and A.S. Funding was acquired by F.S., A.S., and S.A.S.

\section*{DATA AND CODE AVAILABILITY}
The data, simulation and analysis tools/code used in this manuscript can be found in Reference~\cite{Cont_SR_data}. 

\ifpralook\else

\section*{Competing interests}
The authors declare no competing interests.

\section*{Additional information}
Correspondence and requests for materials should be addressed to F.S.

\fi


\section*{Supplementary Information}

\textbf{SR output power.}
Superradiant emission emerges when the collective gain overcomes the decoherence in the system, $\gamma_\mathrm{eff}$ , which is dominated here by the Doppler linewidth $\gamma_\mathrm{D}$. A steady-state collective dipole can then form above the critical intracavity atom number, see Eq.~\eqref{eq:NC}, where the expression has been written consistently with Eq.~\eqref{eq:Nsr}. As shown in Fig.~\ref{fig:Figure_1}d, the SR output power scales linearly with $N_\mathrm{SR}$ for $N_\mathrm{SR} > N_\mathrm{crit}$. Since both cavity mirrors have identical coatings, light of identical intensity is emitted in both directions, and the detected power at one output port is half the total emitted power. Consequently, all mathematical expressions for the measured output power contain a factor $1/2$. We first describe the measured power by the linear approximation~\cite{tangProspectsSuperradiantLaser2022}
\begin{equation} 
P_{\mathrm{out}} = \frac{1}{2}b_{\mathrm{lin}}(N_\mathrm{SR} - N_\mathrm{crit}), 
\label{eq:Power_out_lin} 
\end{equation} 
which is fitted to all data points with $P_\mathrm{out} > 1\,\mathrm{nW}$. This restriction excludes the near-threshold regime, where the power is not strictly linear, and yields the slope $b_\mathrm{lin}$ and experimental threshold $N_\mathrm{crit}^\mathrm{exp}$ for each value of $\gamma_\mathrm{eff}$.
The Doppler-linewidth dependence of the measured slopes is described by the phenomenological model
\begin{equation} 
P_{\mathrm{out}} = \frac{1}{2}\frac{hc}{\lambda}\ b_P \left(\frac{\gamma_T}{\gamma_\mathrm{D}}\right)^{\alpha} \frac{N_\mathrm{SR} - N_\mathrm{crit}}{\tau_T}, 
\label{eq:Power_out} 
\end{equation} 
valid for $\gamma_\mathrm{D} \gg \gamma_T$ and $N_\mathrm{SR} > N_\mathrm{crit}$, where $\tau_T = \sqrt{\pi}w_0/\bar{v}_x$ is the Gaussian-averaged transit time [Eq.~\eqref{eq:tau}]. The factor $b_P(\gamma_T/\gamma_\mathrm{D})^\alpha$ captures the suppression of collective emission by Doppler broadening. Fitting Eq.~\eqref{eq:Power_out} to the measured slopes from Fig.~\ref{fig:Figure_1}d and Ext. Data  Fig.~\ref{fig:Figure_6_methods}b yields $b_P = 1.27 (2) $ and $\alpha = 0.502 (7)$ with $\chi^2_r = 0.69$. 
Accordingly, the effective emission factor $b_P(\gamma_T/\gamma_\mathrm{D})^\alpha$ decreases from approximately $0.75$ to $0.35$ as $\gamma_\mathrm{D}$ is increased from $1$ to $4$\,MHz. Energy conservation requires this factor to remain below unity, which is satisfied for $\gamma_\mathrm{D} \gtrsim 600$\,kHz and hence throughout all experimental conditions considered here.
Our analytical model is derived from the same Hamiltonian used for the NMS description~\cite{famaCollimationDenseAtomic2026}, with the addition of a coherent drive. It predicts the same linear scaling as Eq.~\eqref{eq:Power_out} but yields a Doppler-independent emission factor corresponding to $50\%$ effective participation atom number of in the collective emission process. Numerical simulations in Ref.~\citenum{Rugged_mHz_Linewidth_Haonan_2020} predict that up to $70\%$ of the atoms can participate in the emission, which is close to the largest values we measure. A quantitative microscopic description of the observed Doppler dependence of the emission factor is beyond the scope of this work.

Now, we use Eq.~\eqref{eq:Power_out_lin} to extract $N_\mathrm{crit}^\mathrm{exp}$ from the data (an example can be seen in the inset of Fig.~\ref{fig:Figure_1}d for $\gamma_\mathrm{eff}/(2\pi)=1.47(4)$\,MHz where we retrieve $N_\mathrm{crit}^\mathrm{exp} = 2.53 (5)\times10^4$\,atoms, prediction $N_\mathrm{crit}^\mathrm{th} = 3.9\times10^4$\,atoms) in Fig.~\ref{fig:Figure_1}d and Ext. Data Fig.~\ref{fig:Figure_6_methods}b. Fitting these thresholds with
\begin{equation}
    N_\mathrm{crit}^\mathrm{exp}(\gamma_\mathrm{eff})
    = a\frac{\kappa\gamma_\mathrm{eff}}{2g_0^2}
\end{equation}
yields $a = 0.95(1)$, in close agreement with the theoretical prediction $a=1$. The fitted thresholds are systematically lower by approximately $5\%$. This offset can be explained because the output power initially follows the classic $N^2$ superradiant scaling directly above threshold before crossing over to the linear regime~\cite{tangProspectsSuperradiantLaser2022}. A linear fit restricted to higher power therefore extrapolates to $P_\mathrm{out}=1$\,nW at a slightly underestimated threshold atom number. We note that for low Doppler broadening, the difference between theory and experimental measurement is larger than for larger broadening.

\textbf{Cavity pulling model.}
The inset of Fig.~\ref{fig:figure2}a shows the cavity-pulling suppression $\rho$ extracted from four data sets with different $\gamma_\mathrm{eff}$. To test the validity of Eq.~\eqref{eq:rho} (black line), we fit the data with a single-parameter scaling model,
\begin{equation}
    \rho
    = \frac{a\,\gamma_\mathrm{eff}}{a\,\gamma_\mathrm{eff}+\kappa},
\end{equation}
where $a$ is a dimensionless scaling factor applied to $\gamma_\mathrm{eff}$, the quantity with the largest uncertainty. The fit yields $a = 1.09(9)$, consistent with unity. Although Eq.~\eqref{eq:rho} is a simplified description of the experiment, it reproduces the data remarkably well.

\noindent\textbf{Superradiant laser linewidth.}
\label{Sec:analytical_linewidth}
Our experiment operates in the regime where neither the good-cavity ($\kappa \ll \gamma_\mathrm{D}$) nor the bad-cavity ($\kappa \gg \gamma_\mathrm{D}$) adiabatic elimination fully applies. We describe the system using an analytical model that is based on previous work presented in Refs.~\citenum{LaburtheTolra2026,laburthe-tolraCorrelationsLinewidthAtomic2023} that captures the dominant noise sources and reproduces the overall trends observed experimentally. We note that this analytical model is a simplification and cannot predict the linewidth near transition points, for example between steady-state superradiance and the multi-component superradiant regime (see Ext. Data Fig.~\ref{fig:Figure_6_methods}a). 
The relevant dephasing rate in Refs.~\citenum{LaburtheTolra2026,laburthe-tolraCorrelationsLinewidthAtomic2023} is $\gamma_R$, which represents the rate at which atoms leave the cavity and are replaced by new excited atoms with a random phase, modelled as independent quantum noise. In our continuous atomic beam, $\gamma_R$ is set by the transit time through the cavity mode [Eq.~\eqref{eq:tau}, Methods], depending only on the mean atomic velocity and the cavity waist, $\gamma_R \simeq \gamma_T = 2\pi\times 360\,\mathrm{kHz}$. Compared to this previous work, we also included Doppler broadening, the velocity spread along the cavity direction, in our analytical model. 

Starting with the model, we first consider the intracavity photon number as
\begin{equation}
    N_\nu = \frac{1}{2\kappa}(N_\mathrm{SR} - N_\mathrm{crit})\,\gamma_T,
    \label{eq:photon_number_cont}
\end{equation}
where we neglect transmission losses of the cavity mirrors and vacuum windows. The threshold atom number $N_\mathrm{crit}$ [Eq.~\eqref{eq:NC}, main text] is set by Doppler broadening $\gamma_\mathrm{D} \geq 2\pi\times 1\,\mathrm{MHz} > \gamma_T$ leading to an effective broadening of $\gamma_\mathrm{eff} \geq 2\pi\times 1.4\,\mathrm{MHz}$. The laser linewidth receives two independent contributions~\cite{LaburtheTolra2026,laburthe-tolraCorrelationsLinewidthAtomic2023}. 

The first contribution is \emph{atomic coherence noise}. There are two dominant mechanisms that give rise to atomic coherence noise. The first is replenishment-induced phase noise, arising from the stochastic reset or replacement of emitters as the ensemble is maintained in operation. The second mechanism is Doppler dephasing: atoms with different velocity components along the cavity axis accumulate phase at different rates during their transit through the cavity mode. Together, these two mechanisms contribute phase noise to the collective emission, modeled by the effective decoherence rate $\gamma_\mathrm{eff}$, and give rise to the atomic coherence-noise contribution 
\begin{equation}
    \gamma_1 = \frac{2g_0^2\gamma_\mathrm{eff}}{4(\kappa+\gamma_\mathrm{eff})^2} \frac{N_\mathrm{SR}}{N_\nu}.
    \label{eq:linewidth_atomic_cont}
\end{equation}

The second contribution to the laser linewidth originates from \emph{photon shot noise}, which gives rise to
\begin{equation}
    \gamma_2 = \frac{\kappa}{4N_\nu}\frac{\gamma_\mathrm{eff}^2}{(\kappa+\gamma_\mathrm{eff})^2}=\frac{\kappa}{4N_\nu}\rho^2.
    \label{eq:linewidth_photon_cont}
\end{equation}
In the good-cavity limit $\rho \approx 1$, $\gamma_2$ recovers the standard Schawlow-Townes linewidth, whereas in the bad-cavity limit $\kappa \gg \gamma_{\mathrm{eff}}$, it becomes $\gamma_2 \simeq \gamma_{\mathrm{eff}}^2/(4N_\nu\kappa)$, corresponding to a suppression of the classical value $\kappa/(4N_\nu)$ by the factor $(\gamma_{\mathrm{eff}}/\kappa)^2 \ll 1$.

The total SR laser linewidth can be rewritten as
\begin{equation}
    \gamma_\mathrm{SR} = \gamma_1 + \gamma_2=\frac{\kappa}{4N_\nu}\rho^2 \left( 1+\frac{2g_0^2N_\mathrm{SR}}{\kappa\gamma_\mathrm{eff}} \right).
    \label{eq:linewidth_total}
\end{equation}
In the limit $N_\mathrm{SR} \gg N_\mathrm{crit}$, the two terms behave fundamentally differently. Since $N_\nu \propto (N_\mathrm{SR}-N_\mathrm{crit})$, the photon noise contribution scales as $\gamma_2 \propto 1/N_\mathrm{SR} \to 0$. The atomic coherence term, however, saturates to a constant because $N_\nu \propto N_\mathrm{SR}$.
This asymptotic floor is set entirely by the single-atom coupling $g_0$, Doppler dephasing $\gamma_\mathrm{eff}$, and transit rate $\gamma_\mathrm{T}$, and is independent of the intracavity photon number. The total linewidth therefore approaches a constant at large $N_\mathrm{SR}$, dominated by atomic coherence noise, as visible in Fig.~2b of the main text, where the data flatten for $N_\mathrm{SR} \gtrsim 1\times10^5$.

\textbf{Fundamental superradiant linewidth limit.}
Deep in the bad-cavity limit ($\kappa \gg \gamma_\mathrm{eff}$) and far above threshold ($N_\mathrm{SR} \gg N_\mathrm{crit}$), the atomic coherence noise dominates and the linewidth approaches 
\begin{equation}
\gamma_{\mathrm{SR}}^{\text{Asymptotic}}=\frac{g_0^2\gamma_\mathrm{eff}}{\kappa\gamma_T}.
\end{equation}
In our parameter regime, $\gamma_{\mathrm{eff}}>4\gamma_T$, this asymptotic limit lies above the Purcell linewidth
\begin{equation}
\gamma_\mathrm{SR}^{\mathrm{Purcell}} = \frac{4g_0^2}{\kappa},
    \label{eq:purcell}
\end{equation}
which is independent of the intracavity photon number, gain broadening, and the exact atom number and is set solely by the single-atom coupling $g_0$ and the cavity linewidth $\kappa$~\cite{Prospects_for_a_Millihertz_Linewidth_Meiser}. For our parameters $\gamma_\mathrm{SR}^{\mathrm{Purcell}} = 2\pi\times75\,\mathrm{Hz}$. Under our current operating conditions, the measured linewidth at $N_\mathrm{SR} \approx 1.57(8) \times 10^5$ is $320 (40)\,\mathrm{Hz}$, lying a factor of ${\approx}\,4.3$ above the Purcell floor. Our analytical model [Eq.~\eqref{eq:linewidth_total}] predicts an intrinsic SR linewidth of $\sim2\pi\times90\,\mathrm{Hz}$ at this atom number, lying between the Purcell floor and the measured value. The remaining gap between the prediction and the measurement is consistent with the finite probe laser linewidth and technical noise (see Methods and SI), leaving open the possibility that the intrinsic SR linewidth is already near the Purcell floor.

In the crossover regime ($\rho<1$ and $\Gamma_\mathrm{coll}\gtrsim\kappa$), linewidths \emph{below} the Purcell floor are in principle accessible. The Schawlow-Townes-like suppression factor $(\gamma_\mathrm{eff}/\kappa)^2 \ll 1$ in combination with a large intracavity photon number $N_\nu$ can push the linewidth $\Delta\nu_\mathrm{SR}$ below $\gamma_\mathrm{SR}^{\mathrm{Purcell}}/(2\pi)$, as shown analytically and in numerical simulations for the $^1\!S_0 \to {}^3\!P_1$ transition in Refs.~\citenum{tieriTheoryCrossoverLasing2017,debnathLasingSuperradiantCrossover2018}.
In our system this regime would become visible if $\gamma_\mathrm{eff}$ were reduced towards $\gamma_\mathrm{T}$, accessible in principle by better transverse cooling of the atomic beam, and if the linewidth measurement were not limited by technical noise.

However, this improvement has a hard limit: once $\gamma_\mathrm{eff} \to \gamma_T$, transit-time broadening becomes the dominant dephasing mechanism and further cooling yields no benefit. Increasing $\kappa$ through cavity design is therefore the more direct route for achieving smaller intrinsic linewidth, since it simultaneously lowers the Purcell floor as $4g_0^2/\kappa$, improves cavity pulling suppression as $\gamma_\mathrm{eff}/(\gamma_\mathrm{eff}+\kappa)$, and carries no fundamental lower limit. 

Combining an optimized cavity linewidth $\kappa$, improved transverse cooling, and higher atomic flux through the cavity mode, an intrinsic SR linewidth below $10\,\mathrm{Hz}$ should be achievable in future realizations of strontium hot beam SR lasers.

\begin{figure*}[t]
\centering
\includegraphics[width=\textwidth]{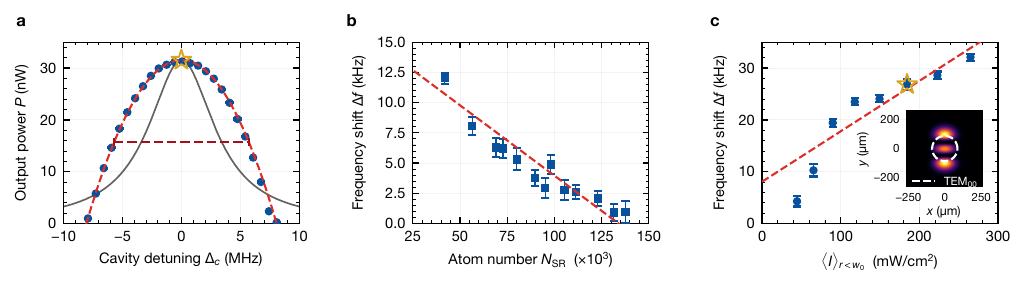}
\caption{
\textbf{Frequency-shift characterization of the superradiant laser.}
\textbf{(a)}~SR output power as a function of cavity detuning $\Delta_c$ (blue points). The dashed red curve is a parabolic fit [Eq.~\eqref{eq:parabol}], yielding a peak power of $31.5 (1)$\,nW and an effective FWHM of $11.40 (2)$\,MHz. The solid black curve shows the bare cavity transmission, with a FWHM of $\kappa/2\pi = 6.79 (2)$\,MHz, for comparison. 
\textbf{(b)}~Superradiant emission frequency $\Delta f_\mathrm{SR}$ as a function of intracavity atom number $N_\mathrm{SR}$, measured via heterodyne beat detection against the Spec laser. The red dashed line is a linear fit, $f_\mathrm{SR} = \chi N_\mathrm{SR} + b$, yielding $\chi = -0.116(6)$\,Hz/atom. 
\textbf{(c)}~SR frequency $\Delta f_\mathrm{SR}$ as a function of the average intracavity intensity $\langle I \rangle_{r<w_0}$ of the TEM$_{02}$ cavity-lock light, measured via heterodyne beat detection against the Spec laser. The red dashed line is a linear fit, $\Delta f_\mathrm{SR} = \alpha\langle I \rangle + \mathrm{offset}$, yielding an effective lock-power-to-SR-frequency coefficient $\alpha = 97(19)$\,Hz\,(mW/cm$^{2}$)$^{-1}$. Inset: TEM$_{02}$ intensity profile. The dashed circle indicates the TEM$_{00}$ mode area, $r < w_0 = 87\,\mu$m. Stars indicate data taken at our reference operating point.
Error bars: (a), as in Fig.~\ref{fig:Figure_1}d; (b, c), as in Fig.~\ref{fig:figure2}.
}
\label{fig:Figure_4_methods}
\end{figure*}

\textbf{Lasing power dependency on cavity detuning.}
During the cavity-pulling measurement, we observe that the SR output power depends on the cavity detuning (for fixed $\gamma_\mathrm{eff}$ and $N_\mathrm{SR}$). We use this dependence experimentally to identify the optimal cavity-lock operating point. We measure the cavity-detuning-dependent SR output power by recording $P$ as a function of cavity detuning $\Delta_c$ (Ext. Data Fig.~\ref{fig:Figure_4_methods}a). The output power depends on both the cavity transmission and the collective atomic gain. The measured profile is described phenomenologically by
\begin{equation}
    P(\Delta_c) = P_0 - a\,(\Delta_c - \Delta_0)^2,
    \qquad
    \mathrm{FWHM}_\mathrm{eff} = \sqrt{\frac{2P_0}{a}},
    \label{eq:parabol}
\end{equation}
where the fit is used only over the measured range with $P(\Delta_c)>0$. The centre of the parabolic fit defines the optimal operating point for the cavity lock. The effective FWHM of the measured power profile is $\mathrm{FWHM}_\mathrm{eff} = 11.40 (2)$\,MHz, broader than the intensity FWHM of the bare cavity transmission, $\kappa/2\pi = 6.79(2)$\,MHz.
We further examine this broadening using the cavity-detuning-dependent critical atom number for lasing obtained from our simple analytical model, which is also used for the linewidth estimate,
\begin{equation}
    N_{\mathrm{crit}}(\Delta_c) =\frac{\gamma_{\mathrm{eff}}}{g_0^2\kappa}
    \left(
    2\Delta_c^2(1-\rho)^2+\frac{\kappa^2}{2}
    \right).
    \label{eq:Ncrit_Dc}
\end{equation}
The SR output vanishes when $N_\mathrm{SR}<N_\mathrm{crit}(\Delta_c)$. For $N_\mathrm{SR}=1.5 \times 10^5$, $\gamma_\mathrm{eff}=2\pi\times1.47(4)$\,MHz, and $\rho=0.165$, Eq.~\eqref{eq:Ncrit_Dc} predicts a threshold detuning of $\Delta_\mathrm{c}^\mathrm{th}=2\pi\times 6.96(10)$\,MHz, below the measured value of $2\pi\times 8.06(1)$\,MHz. At this relatively small effective broadening, the observed lasing range is therefore wider than predicted by the simplified model. We believe this discrepancy can be explained by the systematically lower atom-number threshold observed in Fig.~\ref{fig:Figure_1}, as well as by atom-number fluctuations that broaden the measured detuning-dependent power envelope. Whether a comparable discrepancy persists at larger $\gamma_\mathrm{eff}$ remains to be investigated.

\textbf{Long-term stability, systematic frequency shifts.}
In an ideal superradiant laser the emission frequency is set by the atomic transition, as cavity pulling is negligible ($\kappa \gg \gamma_\mathrm{eff}$), and the fundamental stability limit is imposed by phase diffusion. Newly arriving excited state atoms add quantum noise to the phase of the collective atomic dipole while they synchronize with it. In the regime where this diffusion dominates, this leads to a random walk of the SR laser phase~\cite{kazakovUltimateStabilityActive2022,Rubiola_chart}
\begin{equation}
    S_\nu^\mathrm{phase} 
    = \frac{\Delta\nu_\mathrm{SR}}{\pi}.
    \label{eq:Snu_phase}
\end{equation}
The white frequency modulation noise gives an Allan deviation scaling~\cite{Rubiola_chart} of $\tau^{-1/2}$:
\begin{equation}
    \sigma_y^\mathrm{phase}(\tau) 
    = \sqrt{\frac{S_\nu^\mathrm{phase}}{2\,\nu_0^2\,\tau}}
    = \frac{1}{2\pi\,\nu_0}\sqrt{\frac{\gamma_\mathrm{SR}^{\mathrm{Purcell}}}{\tau}}
    \approx \frac{8.0\times10^{-15}}{\sqrt{\tau/\mathrm{s}}}.
    \label{eq:sigma_phase}
\end{equation}
The measured instability in the self-locked case (where the long-term stability is defined by the atomic transition), $\sigma_y(\tau) = 2.0(1)\times10^{-12}\,(\tau/\mathrm{s})^{-0.29(5)}$ (Fig.~\ref{fig:Figure_3}), lies more than two orders of magnitude above the Purcell limit at $\tau = 1$\,s, indicating that technical noise sources dominate the stability at all measured timescales.
The Allan deviation data reveal three distinct regimes, classified according to the noise spectra of Ref.~\citenum{Rubiola_chart}. At short timescales $\tau \lesssim 50$\,ms, the SR laser operates free-running above the cavity PDH lock bandwidth, and the upper bound on the intrinsic linewidth is $\Delta\nu_\mathrm{meas} = 320(40)$\,Hz. 
The $\beta$-separation linewidth $\Gamma_{\beta,\mathrm{self}} = 2720(82)$\,Hz integrates the frequency noise PSD above $\sim$10\,Hz and captures the additional contribution from the SR cavity PDH lock (Ext. Data Fig.~\ref{fig:Figure_3_methods}c). The ratio $\Gamma_{\beta,\mathrm{self}}/\nu_0 \approx 6\times10^{-12}$ is consistent in order of magnitude with the observed $\sigma_y(1\,\mathrm{s}) = 2.0\times10^{-12}$, indicating that the 1\,s instability is set by the PDH lock noise rather than by phase diffusion. At the intermediate timescales $0.05\,\mathrm{s} \lesssim \tau \lesssim 4$\,s, the instability is dominated by the residual SR PDH lock noise and shot-to-shot atom-number fluctuations, both of which contribute independently (Fig.~\ref{fig:Figure_3}b). The peak at $\tau \approx 0.2$\,s is the servo bump arising from the PDH lock corner frequency of $\sim$2.5\,Hz.  
At longer timescales $\tau \gtrsim 4$\,s, the slope flattens towards the observed $\tau^{-0.29}$, consistent with flicker FM noise~\cite{Rubiola_chart}, which could be attributed to slow drifts of the atom number $N_\mathrm{SR}$ due to intensity fluctuations in the state preparation and cooling beams, and to the lock-laser intensity $\langle I\rangle$ drift, neither of which is actively power stabilized. The overall fit exponent $\tau^{-0.29}$ over the full range $0.2$\,s to $400$\,s reflects the mixture of these white FM and flicker FM contributions and does not correspond to a single noise process. 

\newcommand{\ExtDataTable}{

\gdef\tablename{Extended Data Table}
\begin{table}[t]
\centering
\caption{
    \textbf{Dominant frequency-shift contributions} at the reference operating point ($N_{\rm SR} = 1.53(2) \times 10^5, \gamma_{\rm eff}/(2\pi) = 1.47(4)\,$MHz). 
    $\sigma_f$ is the estimated frequency instability from a $\pm5\%$ fluctuation of the respective control parameter. The lock-light-induced shift is independent of the atom number and is evaluated around $\langle I\rangle\approx190$\,mW/cm$^2$.
    }
\label{tab:freq_shifts}
\begin{tabular}{llcc}
\hline
Source & Mechanism & Coefficient & $\sigma_f$ (Hz) \\
\hline
Atom number $N_\mathrm{SR}$
    & \makecell[l]{Atom-number-dependent\\frequency shift}
    & $-0.116 (6)\ \mathrm{Hz/atom}$
    & $890 (50)$ \\
Lock-laser intensity
    &  \makecell[l]{Effective lock-power-induced\\ frequency shift}
    & $97(19)\ \mathrm{Hz/(mW/cm^2)}$
    & $920 (180)$ \\
\hline
\end{tabular}
\end{table}
}

\ifdoublecolumnlook
    \ExtDataTable
\fi

\textbf{Atom-number–dependent frequency shift.}
We observe that the SR laser frequency depends on the intracavity atom number $N_\mathrm{SR}$ (Ext. Data Fig.~\ref{fig:Figure_4_methods}b), and therefore atom number fluctuations lead to frequency fluctuations. A simplified analytical model, which neglects Doppler broadening and does not include the full cavity-detuning dependence, predicts an atom-number-dependent cavity-pulling factor

\begin{equation}
    \rho(N_\mathrm{SR})=\frac{1}{1+\Gamma_\mathrm{coll}\kappa\tau_T^2/32},
    \label{eq:pulling_N}
\end{equation}
where $\Gamma_\mathrm{coll}=
2g_0^2 N_\mathrm{SR}/\kappa$ is the collectively enhanced emission rate and $\tau_T$ is the transit time.
The atom number dependence of the pulling factor leads to frequency fluctuations if the cavity is detuned from resonance. However, our analytical model neglects Doppler-broadening and other effects and therefore we cannot uniquely assign the measured dependence to this mechanism. A detailed microscopic description is left for future work.
Experimentally, we observe an approximately linear dependence,
\begin{equation}
    \Delta f_\mathrm{SR} = \chi N_\mathrm{SR},
\end{equation}
with $\chi=-0.116(6)$\,Hz/atom. Small variations of the fitted slope are observed between $\Delta f(N_{\rm SR})$ scans measured days apart. These variations might originate from residual changes in the cavity detuning.
As an example, at our reference operating point of $N_\mathrm{SR}=1.53 \times 10^5$ atoms a $5\%$ atom-number fluctuation corresponds to $\Delta N_\mathrm{SR}=7\,650$, leading to a frequency shift of $890(50)$\,Hz. Here, the uncertainty includes the scatter of the measured frequency dependence beyond the statistical uncertainty of the linear fit. 

\textbf{Lock-light-intensity-dependent frequency shift.}
Another important contribution to the long-term frequency stability is the dependence of the SR laser frequency on the cavity-lock-light intensity. The cavity resonance frequency is stabilized using a PDH locking scheme. The lock light, with frequency $\omega_\mathrm{lock}$, is red detuned by $2650$\,MHz from the ${}^1\!S_0\rightarrow{}^3\!P_1$ transition to avoid directly driving the SR transition. However, this detuning is not at a magic wavelength, so the lock laser induces a light shift that must be characterized and ideally controlled. Additionally, we expect that a change in intensity on the PDH photodiode could shift the lock point. The lock light is coupled into a TEM$_{02}$ cavity mode, which has a convenient detuning for cavity locking and reduced spatial overlap with the TEM$_{00}$ SR mode. This reduces the light shift experienced by the atoms. We further reduce the light shift by using less than $1\,\mu$W of lock power before the cavity.

The lock laser has linear polarization orthogonal to the quantization axis, which is defined by the magnetic field $\mathrm{B}$ (Ext. Data Fig.~\ref{fig:Figure_1_methods}). The intracavity power is related to the power before the cavity by
\begin{equation}
    P_{\mathrm{intra}} = \eta_{\mathrm{cav}}\,\eta_{\mathrm{coup}}
      \,\frac{F}{\pi}\,P_{\mathrm{exp}},
    \label{eq:P_intra}
\end{equation}
where $F=828$ is the cavity finesse, $\eta_\mathrm{cav}=0.50$ is the estimated transmission of optical elements between the last accessible point in the beam path at which $P_{\mathrm{exp}}$ is measured and the cavity, and $\eta_\mathrm{coup}=0.25$ is the coupling efficiency into the cavity mode. We define the effective lock-light intensity as the uniform spatial average of the TEM$_{02}$ intracavity intensity over the circular area $r<w_0$ corresponding to the TEM$_{00}$ waist. It is given by
\begin{equation}
    \langle I \rangle_{r<w_0}
    = 2P_{\mathrm{intra}}\,\mathcal{I}_{\mathrm{norm}},
    \label{eq:I_avg}
\end{equation}
where $\mathcal{I}_{\mathrm{norm}}$ is the transverse intensity prefactor obtained by averaging the normalized TEM$_{02}$ mode profile over the disk $r<w_0$, with $w_0=87\,\mu\mathrm{m}$. The factor of two accounts for the longitudinally averaged standing-wave intensity in the linear cavity, relative to the circulating travelling-wave power $P_\mathrm{intra}$. For equal intracavity powers, this spatial average is reduced by $68\%$ for the TEM$_{02}$ mode compared to the TEM$_{00}$ mode assuming perfect longitudinal overlap. 
The light shift of the SR frequency is therefore expected to be linear in the lock-light intensity,
\begin{equation}
    \Delta f_{\mathrm{shift}}
    = -\Delta\alpha(\omega_\mathrm{lock}) \cdot \langle I \rangle_{r<w_0},
    \label{eq:lightshift}
\end{equation}
where $\Delta\alpha(\omega_\mathrm{lock})$ is the differential polarizability of the lasing states at the lock-laser frequency, accounting for the linear polarization orthogonal to the quantization axis,
\begin{equation}
    \Delta\alpha(\omega_\mathrm{lock})
    = \alpha_{1S_0}(\omega_\mathrm{lock}) - \alpha_{3P_1,m_J=0}(\omega_\mathrm{lock}).
    \label{eq:pol}
\end{equation}
Using $\alpha_{1S_0}(\omega_\mathrm{lock})=-0.95212\,\mathrm{Hz}/(\mathrm{mW}/\mathrm{cm}^2)$ and $\alpha_{3P_1,m_J=0}(\omega_\mathrm{lock})=0.6798\,\mathrm{Hz}/(\mathrm{mW}/\mathrm{cm}^2)$ gives $\Delta\alpha(\omega_\mathrm{lock})=-1.6319\,\mathrm{Hz}/(\mathrm{mW}/\mathrm{cm}^2)$ and hence a predicted positive transition-frequency shift of $1.6319\,\mathrm{Hz}/(\mathrm{mW}/\mathrm{cm}^2)$ (Ref.~\citenum{urech2023single}).

We measure the lock-light-induced frequency shift for several lock light powers using the SR-Spec beat signal, which delivers $\sim5$\,kHz precision (Ext. Data Fig.~\ref{fig:Figure_4_methods}c). When interpreting these data, note that changes in lock power also affect the PDH error-signal amplitude. In particular, at low powers the photodiode dark current contribution becomes significant and can shift the lock point. We therefore measure an effective lock-power-to-SR-frequency coefficient rather than the pure atomic light-shift coefficient. A linear fit yields
\begin{equation}
    \alpha = 97(19)\ \mathrm{Hz\,(mW\,cm^{-2})^{-1}},
    \qquad
    \chi^2_\mathrm{red} = 22.69.
    \label{eq:alpha_meas}
\end{equation}
The calculated atomic light shift amounts to $\sim2\%$ of $\alpha$, indicating that technical lock-point shifts dominate the observed dependence. At low intensities, $\langle I \rangle<100$\,mW/cm$^2$, where technical contributions are stronger, the fit describes the data poorly. At the operating point, $\langle I \rangle\approx190$\,mW/cm$^2$, the data follow an approximately linear trend. 
For slow, quasi-static $\pm5\%$ fluctuations of the lock-light intensity around the operating point, the corresponding frequency instability is
\begin{equation}
    \sigma_f^{(I)} = \alpha \cdot 0.05\,\langle I \rangle^\mathrm{op}
    = 920 (180)\,\mathrm{Hz},
    \label{eq:lightshift_sensitivity}
\end{equation}
corresponding to a fractional frequency instability of $\sigma_y\approx2\times10^{-12}$ on time scales of minutes. This contribution is comparable to the atom-number-dependent frequency shift of $890(50)$\,Hz from the previous subsection. Both contributions, atom-number and lock-light intensity fluctuations, can explain major parts of the observed long-term frequency stability.

The two frequency-shift mechanisms are summarized in Ext. Data Table~\ref{tab:freq_shifts}. Both contributions scale approximately linearly with their respective control parameters, $N_\mathrm{SR}$ and $\langle I\rangle$, around the operating point. In principle, they can be reduced through improved stabilization: the atom number through cooling-beam power stabilization and more accurate centring of the cavity resonance to the SR output frequency, and the lock light through intensity feedback control. A cavity with higher finesse at a magic frequency for the locking light would reduce the lock-light intensity shift to negligible levels. 

{\bf Frequency noise characterization.}
The frequency noise PSD $S_\nu(f)$ is computed from the ULE-SR beat signal recorded over 1\,s on a digital oscilloscope (R\&S MXO44) via a Hilbert transform of the bandpass-filtered beat waveform, using Welch's method 
($n_\mathrm{seg} = 8$, Hann window, 50\% overlap). The dominant noise contribution arises from the finite SNR of 
the cavity length PDH lock (SNR\,$\approx$\,500-1000). In combination with the inverse cavity-pulling suppression factor $1/\rho \approx 6$ ($\gamma_\mathrm{eff}/2\pi = 1.47(4)$\,MHz, $\rho = 0.165 (3)$), this limits the SR frequency noise to
\begin{equation}
    \Delta\nu_\mathrm{SR} \sim 
    \frac{\kappa \rho}{\mathrm{SNR}}
    = \frac{6.79\,\mathrm{MHz}}{500 \times 6}
    \approx 2.3\,\mathrm{kHz},
    \label{eq:SR_noise}
\end{equation}
consistent with the $\beta$-separation linewidth~\cite{DiDomenico2010}
\begin{equation}
    \Delta\nu_\beta = \sqrt{8\ln 2
        \int_{S_\nu > \beta f} S_\nu\,\mathrm{d}f},
    \qquad \beta = \frac{8\ln 2}{\pi^2} \approx 0.56,
    \label{eq:beta_linewidth}
\end{equation}
yielding $\Delta\nu_{\beta_{ULE}}= 2172 (70)$\,Hz under direct ULE laser lock of the SR cavity length and $\Delta\nu_{\beta_{self}} = 2720 (82)$\,Hz under self-lock (Ext. Data Fig.~\ref{fig:Figure_3_methods}c). Uncertainties are estimated by bootstrapping over $n = 200$ Welch-segment resamples. The approximately $550$\,Hz increase under self-lock is consistent with residual ringing from the finite cavity-pulling suppression: the feedback loop has an effective gain of $\rho = 0.165(3)$ at the SR output, adding a noise contribution of order $\rho\Delta\nu_{\beta_{ULE}} \approx 360$\,Hz. 

In summary, we note that $\Delta\nu_\mathrm{meas}$ and $\Delta\nu_\beta$ characterize complementary aspects of the emission: $\Delta\nu_\beta = 2172 (70)$\,Hz is extracted from the full frequency noise PSD over 1\,s and reflects the laser as seen by an external user including all technical noise contributions as cavity lock-induced noise, whereas $\Delta\nu_\mathrm{meas} = 320 (40)$\,Hz is extracted at $T = 11$\,ms, below the lock bandwidth, and reflects an upper bound of the intrinsic SR emission linewidth. Reducing the cavity locking light linewidth to $\lesssim 50$\,kHz at similar SNR would bring $\Delta\nu_\mathrm{SR}$ to the sub-100\,Hz level, at which point the lock would no longer be limiting.

\subsection*{Analysis of Amplitude Modulations}
\label{sec:phase_transition}
In this section, we first discuss the experimentally extracted parameters and then describe the theoretical microscopic model used for comparison with the experimental data.

{\bf Single-frequency to multi-component lasing.}
We first discuss the atom number scan with fixed  $\gamma_\mathrm{eff}\approx4.5$\,MHz (see Fig.~\ref{fig:Figure_4}c and d). The emission threshold is extracted from the mean output power $\langle P \rangle$ shown in Ext. Data Fig.~\ref{fig:Figure_6_methods}b. Above threshold, $\langle P \rangle$ scales approximately linearly with $N_\mathrm{SR}$ [Eq.~\eqref{eq:Power_out_lin}], and the threshold atom number $N_\mathrm{crit}$ is determined as the zero crossing of a weighted linear fit to $P(N)$, yielding $N_\mathrm{crit}^\mathrm{exp} = 1.13(1) \times 10^5$, which is in good agreement with the theoretical prediction $N_\mathrm{crit}^\mathrm{theo} = 1.18\times 10^5$ (grey dashed line in Fig.~\ref{fig:Figure_4}c,d and Ext. Data Fig.~\ref{fig:Figure_6_methods}b).

The emergence of coherent amplitude oscillations is identified as the lowest scan point at which sidebands are clearly resolved above the noise floor in either the beat-note spectrum or the intensity noise PSD. Both observables yield consistent results. The uncertainty on the boundary is taken as the step size between adjacent scan points: $\Delta\gamma_\mathrm{eff}/(2\pi) = 0.1$\,MHz for the $\gamma_\mathrm{eff}$ scan 
and around $\Delta N_\mathrm{SR} = 5000$ atoms for the $N_\mathrm{SR}$ scan at the given transition point. This gives $\gamma_\mathrm{eff}^\mathrm{MCSR}/(2\pi) = 1.8 (1)$\,MHz (Fig.~\ref{fig:Figure_4}a, vertical red dash-dotted line) and $N^\mathrm{MCSR}_\mathrm{SR} = 1.24(5) \times 10^5$ (Fig.~\ref{fig:Figure_4}c, vertical red dash-dotted line).

The amplitude modulation frequency $f_\mathrm{mod}$ is extracted in two different ways. First, using the intensity noise PSD spectrum itself, where the DC component is removed, and second, using the frequency beat spectrum, where the carrier frequency $f_\mathrm{carrier}$ and each sideband frequency $f_\mathrm{sideband}$ are fitted independently (using a Lorentzian distribution). The modulation frequency is $f_\mathrm{mod} = |f_\mathrm{sideband} - f_\mathrm{carrier}|$, with the uncertainty propagated from the individual fit errors. From the intensity PSD, $f_\mathrm{mod}$ is extracted directly as the peak frequency of a Lorentzian fit to the intensity noise PSD spectrum.

The two methods yield consistent values within their respective uncertainties, which is a non-trivial result: a pure phase modulation of the SR field would produce sidebands in the beat spectrum but leave the intensity noise PSD unaffected. The agreement between the two observables therefore directly confirms that our observation is driven by an oscillation of the collective dipole \textit{amplitude}.

For the $\gamma_\mathrm{D}$ scan (Fig.~\ref{fig:Figure_4}a, b and Ext. Data Fig.~\ref{fig:Figure_6_methods}a), at the well-resolved operating point $\gamma_\mathrm{eff}/(2\pi) = 3.0$\,MHz, $N_\mathrm{SR} = 1.57 \times 10^5$, the beat spectrum yields $f_\mathrm{mod,1}^\mathrm{beat} = 0.94 (4)$\,MHz and $f_\mathrm{mod,2}^\mathrm{beat} = 1.86 (4)$\,MHz, and the intensity PSD yields $f_\mathrm{mod,1}^\mathrm{PSD} = 0.93 (3)$\,MHz and $f_\mathrm{mod,2}^\mathrm{PSD} = 1.86 (3)$\,MHz. Both are in agreement with the theoretical predictions $f_\mathrm{mod,1}^\mathrm{theo} = 0.93(1)$\,MHz for this operating point. We note that in the beat spectrum an additional feature is visible on the negative-frequency side at approximately $-f_\mathrm{mod,2}$, which we attribute to a beat with a spurious sideband of the ULE reference laser or to a background artefact; this feature is absent on the positive-frequency side and has no counterpart in the intensity noise PSD, confirming it is not of SR origin. 

For the $N_\mathrm{SR}$ scan at $\gamma_\mathrm{eff}/(2\pi) = 4.5$\,MHz and $N_\mathrm{SR} = 1.54(5) \times 10^5$, the beat spectrum yields $f_\mathrm{mod,1}^\mathrm{beat} = 0.70 (3)$\,MHz and the intensity PSD yields $f_\mathrm{mod,1}^\mathrm{PSD} = 0.71 (2)$\,MHz, both close to the theoretical prediction $f_\mathrm{mod}^\mathrm{theo} = 0.53(1)$\,MHz. The lower modulation frequency compared to the $\gamma_\mathrm{eff}$ scan reflects the lower gain above threshold.

{\bf Theoretical Modelling.}
We describe superradiant lasing from an atomic beam using the microscopic model developed in Refs.~\citenum{Rugged_mHz_Linewidth_Haonan_2020,SR_thermal_atomic_beam_Jager_Simon}. The dynamics of the semiclassical cavity field $\alpha$ and the spin densities
$s({\bf x},{\bf p},t)=\sum_j s_j\delta({\bf x}-{\bf x}_j)\delta({\bf p}-{\bf p}_j)$ and
$s^z({\bf x},{\bf p},t)=\sum_j s_j^z\delta({\bf x}-{\bf x}_j)\delta({\bf p}-{\bf p}_j)$,
where $s_j$ and $s_j^z$ represent the corresponding operators $\hat{\sigma}_j^-$ and $\hat{\sigma}_j^z$ within a discrete truncated Wigner sampling, are governed by the stochastic differential equations~\cite{SR_thermal_atomic_beam_Jager_Simon,famaContinuousCavityQED2024}
\begin{align}
	\frac{d\alpha}{dt} &=-\frac{\kappa}{2}\alpha-ig_0J+\sqrt{\kappa}\mathcal{F}(t),\label{eq:alpha}\\
	\frac{\partial s}{\partial t} &=-\frac{\bf p}{m}\cdot\nabla_{\bf x}s
	+ig_0\eta({\bf x})s^z\alpha,\label{eq:mfs}\\
	\frac{\partial s^z}{\partial t} &=-\frac{\bf p}{m}\cdot\nabla_{\bf x}s^z
	+i2g_0\eta({\bf x})\left(\alpha^*s-s^*\alpha\right).\label{eq:mfz}
\end{align}
We use an effective two-dimensional geometry, with $z$ along the cavity axis and $x$ along the atomic beam. The cavity mode function is
\begin{align}
	\eta({\bf x})=\cos(kz)e^{-x^2/w^2}.\label{eq:eta}
\end{align}
The collective dipole is given by $
	J=\int d{\bf x}\int d{\bf p}\,\eta({\bf x})s({\bf x},{\bf p},t).$ 
We introduced the notation $\nabla_{\bf x}=(\partial_x,\partial_z)^T$. The boundary conditions at the entrance plane ${\bf x}_0=(x_0,z_0)$, with $x_0\rightarrow-\infty$, are $
	s({\bf x}_0,{\bf p})=s_0({\bf x}_0,{\bf p})
	=\sum_j s_j^{(0)}\delta(x-x_j)\delta(p-p_j),$ and $
	s^z({\bf x}_0,{\bf p})=\sum_j s_j^{z,(0)}
	\delta(x-x_j)\delta(p-p_j)=s_0^z({\bf p}).$
The atoms enter in the excited state, $s_j^{z,(0)}=1$, with
$s_j^{(0)}=(s_j^{x,(0)}-is_j^{y,(0)})/2$, where $s_j^{x,(0)}$ and $s_j^{y,(0)}$ are independently sampled from $\pm1$. Cavity shot noise is included through $\mathcal{F}=(\mathcal{F}^x-i\mathcal{F}^y)/2$, where $\mathcal{F}^x$ and $\mathcal{F}^y$ are independent zero-mean noise sources satisfying
$\langle\mathcal{F}^x(t)\mathcal{F}^x(t')\rangle
=\langle\mathcal{F}^y(t)\mathcal{F}^y(t')\rangle
=\delta(t-t')$.

We assume the incoming atomic distribution to be spatially homogeneous, which is valid when the transverse beam width is much larger than the wavelength of the mode function. We can therefore restrict the spatial integration to one cavity wavelength, $\int d{\bf x}=\int_0^\lambda dz\int_{-\infty}^{\infty}dx$ and the momentum integration is $\int d{\bf p}=\int_0^\infty dp_x\int_{-\infty}^{\infty}dp_z$.
The initial longitudinal spin density is modelled as
\begin{align}
	s_0^z({\bf p})
	=\sqrt{\frac{2}{\pi}}
	\frac{N_{\mathrm{SR}}f_x(p_x/m)f_z(p_z/m)}
	{w\lambda m^2}.\label{eq:s0z}
\end{align}
Normalization is chosen such that the mean squared collective coupling matches with $N_{\mathrm{SR}}g_0^2/2$,
\begin{align}
	\frac{N_{\mathrm{SR}}g_0^2}{2}
	=\int d{\bf x}\int d{\bf p}\,
	g_0^2\eta^2({\bf x})s_0^z({\bf p}).
\end{align}
Here, $f_x$ and $f_z$ are normalized velocity distributions along the $x$ and $z$ directions. Along the beam axis, we use
\begin{align}
	f_x(v_x)=\beta_x m v_x
	e^{-\beta_x m v_x^2/2},\label{eq:fx}
\end{align}
which gives $\bar v_x=\sqrt{\pi/(2m\beta_x)}\approx 345\,\mathrm{m/s}$. The transverse velocity distribution is taken to be Lorentzian,
\begin{align}
	f_z(v_z)=\frac{\Delta v_z}{2\pi}
	\frac{1}{v_z^2+\Delta v_z^2/4},\label{eq:fz}
\end{align}
corresponding to a Doppler broadening $\gamma_D=k\Delta v_z.$
The total effective broadening is then calculated with $\gamma_{\mathrm{eff}}
	=\gamma_D+2\pi\times360\,\mathrm{kHz}
	+2\pi\times7.48\,\mathrm{kHz}.$

\begin{figure}[b]
\centering
\includegraphics[width=\singlecolwidth]{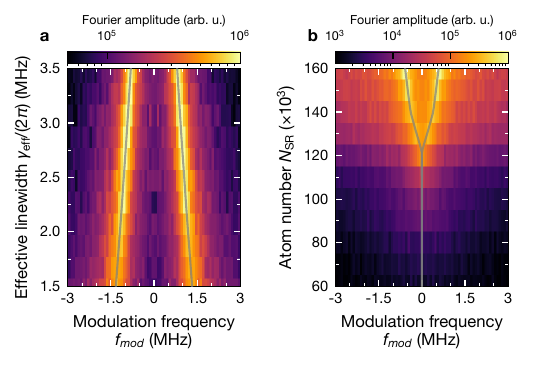}
\caption{\textbf{Theoretical modelling of the amplitude oscillations.} Absolute value of the Fourier transform $\tilde{N}_\nu(\omega)$ [Eq.~\eqref{eq:Ntilde}] as a function of frequency $f_\mathrm{mod}=\omega/(2\pi)$ and for {\bf (a)} varied $\gamma_{\mathrm{eff}}$ and {\bf (b)} varied atom number $N_\mathrm{SR}$. The remaining simulation parameter for {\bf (a)} is $N_{\mathrm{SR}}=1.57 \times 10^5$ and for {\bf (b)} $\gamma_D/(2\pi) = 4.1$\,MHz, resulting in $\gamma_\mathrm{eff}/(2\pi)\approx4.5$\,MHz. The solid lines in {\bf (a)} and {\bf (b)} mark the frequency peaks of a Gaussian fit to $\tilde{N}$. The same solid lines are shown in  Fig.~\ref{fig:Figure_4}a,c.}\label{Fig:Theory}
\end{figure}

{\bf Numerical simulation of the intensity modulation.} We numerically integrate the coupled stochastic equations~\eqref{eq:alpha}-\eqref{eq:mfz} and obtain stochastic trajectories of the intracavity photon number,
\begin{align}
	N_\nu(t)=|\alpha(t)|^2.
\end{align}
The total simulation time is $T=250\,\mu\mathrm{s}$. To characterize the intensity fluctuations, we calculate the Fourier transform of the photon-number fluctuations,
\begin{align}
	\tilde N_\nu(\omega)
	=\int_0^Tdt\,e^{i\omega t}
	\left[N_\nu(t)-\bar N_\nu\right],\label{eq:Ntilde}
\end{align}
where $\bar N_\nu=\frac{1}{T}\int_0^Tdt\,N_\nu(t)$
is the time-averaged photon number.

The results of $\tilde{N}$ obtained from the numerical simulations are visible in Ext. Data Fig.~\ref{Fig:Theory}. In the simulation, following the experimental setting shown in Fig.~4 in the main text, we vary $\gamma_D$ and fix $N_\mathrm{SR}=1.57 \times 10^5$ (Ext. Data Fig.~\ref{Fig:Theory}a) and we fix $\gamma_D/(2\pi)=4.1$\,MHz and vary $N_{\mathrm{SR}}$ (Ext. Data Fig.~\ref{Fig:Theory}b). In addition, we fit to the numerical data a Gaussian model to extract the dominant frequency contribution. The result of this fitting procedure is displayed as solid lines in Fig.~\ref{fig:Figure_4} and Ext. Data Fig.~\ref{Fig:Theory}. These lines represent only the dominant amplitude oscillation frequency which is the first harmonic. Since the underlying microscopic theory is nonlinear in the cavity field, it generally also predicts all higher harmonics. These are, however, progressively suppressed with increasing order and therefore not resolved in the stochastic simulations because of the limited signal-to-noise ratio. 

\ifdefined\processdelayedfloats\processdelayedfloats\fi

\ifdoublecolumnlook
\else
    \newpage
    \ExtDataTable
\fi
\bibliography{main}

@article{SR_thermal_atomic_beam_Jager_Simon,
  title = {Superradiant emission of a thermal atomic beam into an optical cavity},
  author = {J\"ager, Simon B. and Liu, Haonan and Cooper, John and Nicholson, Travis L. and Holland, Murray J.},
  journal = {Phys. Rev. A},
  volume = {104},
  issue = {3},
  pages = {033711},
  numpages = {20},
  year = {2021},
  month = {Sep},
  publisher = {American Physical Society},
  doi = {10.1103/PhysRevA.104.033711},
}

@article{Ultranarrow_Superradiant_Lasing_Zhang_Yuan,
  title = {Ultranarrow Superradiant Lasing by Dark Atom-Photon Dressed States},
  author = {Zhang, Yuan and Shan, Chongxin and M\o{}lmer, Klaus},
  journal = {Phys. Rev. Lett.},
  volume = {126},
  issue = {12},
  pages = {123602},
  numpages = {6},
  year = {2021},
  month = {Mar},
  publisher = {American Physical Society},
  doi = {10.1103/PhysRevLett.126.123602},
}

@article{Prospects_for_a_Millihertz_Linewidth_Meiser,
  title = {Prospects for a Millihertz-Linewidth Laser},
  author = {Meiser, D. and Ye, Jun and Carlson, D. R. and Holland, M. J.},
  journal = {Phys. Rev. Lett.},
  volume = {102},
  issue = {16},
  pages = {163601},
  numpages = {4},
  year = {2009},
  month = {Apr},
  publisher = {American Physical Society},
  doi = {10.1103/PhysRevLett.102.163601},
}

@Article{Chen09,
author="Chen, JingBiao",
title="Active optical clock",
journal="Chinese Science Bulletin",
year="2009",
month="Feb",
day="01",
volume="54",
number="3",
pages="348--352",
issn="1861-9541",
doi="10.1007/s11434-009-0073-y",
}

@article{Thompson1992ObservationNormalMode,
  author  = {R. J. Thompson and G. Rempe and H. J. Kimble},
  title   = {Observation of normal-mode splitting for an atom in an optical cavity},
  journal = {Phys. Rev. Lett.},
  volume  = {68},
  pages   = {1132--1135},
  year    = {1992},
  doi     = {10.1103/PhysRevLett.68.1132},
}

@article{Zhu1990VacuumRabiSplitting,
  author  = {Yifu Zhu and Daniel J. Gauthier and S. E. Morin and Qilin Wu and H. J. Carmichael and T. W. Mossberg},
  title   = {Vacuum Rabi splitting as a feature of linear-dispersion theory: Analysis and experimental observations},
  journal = {Phys. Rev. Lett.},
  volume  = {64},
  pages   = {2499--2502},
  year    = {1990},
  doi     = {10.1103/PhysRevLett.64.2499},
}

@article{Raizen1989NormalModeSplitting,
  author  = {M. G. Raizen and R. J. Thompson and R. J. Brecha and H. J. Kimble and H. J. Carmichael},
  title   = {Normal-mode splitting and linewidth averaging for two-state atoms in an optical cavity},
  journal = {Phys. Rev. Lett.},
  volume  = {63},
  pages   = {240--243},
  year    = {1989},
  doi     = {10.1103/PhysRevLett.63.240},
}

@phdthesis{StellmerPhD2013,
  author  = {Simon Stellmer},
  title   = {Degenerate Quantum Gases of Strontium},
  school  = {University of Innsbruck},
  year    = {2013},
  url     = {https://www.strontiumbec.com/StrontiumLab/Theses/Simon_Stellmer_PhD_thesis.pdf}
}

@article{qiao_ultrastable_2019,
	title = {An ultrastable laser system at 689 nm for cooling and trapping of strontium},
	volume = {125},
	issn = {0946-2171, 1432-0649},
	doi = {10.1007/s00340-019-7328-3},
	number = {11},
	journal = {Appl. Phys. B},
	author = {Qiao, Chang and Tan, C. Z. and Hu, F. C. and Couturier, Luc and Nosske, Ingo and Chen, Peng and Jiang, Y. H. and Zhu, Bing and Weidemüller, Matthias},
	month = nov,
	year = {2019},
	pages = {215},
}

@phdthesis{BeliSilva2025CavityAtom,
  author  = {C. {Beli Silva}},
  title   = {Cavity-atom interactions in a thermal strontium beam: Development of a simple apparatus for quantum metrology},
  school  = {University of Amsterdam},
  year    = {2025},
  url     = {https://dare.uva.nl/id/6c7b4a9b-77b2-48b9-b386-a7ca2655a448}
}

@phdthesis{Fama:25,
  author       = {Fam\`a, Francesca},
  title        = {Waltz of the Atoms: Collective Behaviors in a Sr Stream},
  school       = {University of Amsterdam},
  faculty      = {Faculty of Science (FNWI)},
  institute    = {Institute of Physics (IoP)},
  supervisor   = {Schreck, Florian E.},
  cosupervisor = {Sch{\"a}ffer, Stefan A.},
  year         = {2025},
  date         = {2025-03-14},
  pages        = {157},
  url          = {https://dare.uva.nl/id/2e29368c-d9f5-487f-a9c1-5c0948ff6c60},
}

@article{Schaffer2020,
  title = {Lasing on a narrow transition in a cold thermal strontium ensemble},
  author = {Sch{\"a}ffer, Stefan A. and Tang, Mikkel and Henriksen, Martin R. and J{\o{}}rgensen, Asbj{\o{}}rn A. and Christensen, Bjarke T. R. and Thomsen, Jan W.},
  journal = {Phys. Rev. A},
  volume = {101},
  issue = {1},
  pages = {013819},
  numpages = {9},
  year = {2020},
  month = {Jan},
  publisher = {American Physical Society},
  doi = {10.1103/PhysRevA.101.013819}
}

@article{Laske2019,
  title = {Pulse Delay Time Statistics in a Superradiant Laser with Calcium Atoms},
  author = {Laske, Torben and Winter, Hannes and Hemmerich, Andreas},
  journal = {Phys. Rev. Lett.},
  volume = {123},
  issue = {10},
  pages = {103601},
  numpages = {5},
  year = {2019},
  month = {Sep},
  publisher = {American Physical Society},
  doi = {10.1103/PhysRevLett.123.103601}
}

@article{DiDomenico2010,
author = {Gianni Di Domenico and St\'{e}phane Schilt and Pierre Thomann},
journal = {Appl. Opt.},
number = {25},
pages = {4801--4807},
publisher = {Optica Publishing Group},
title = {Simple approach to the relation between laser frequency noise and laser line shape},
volume = {49},
month = {Sep},
year = {2010},
doi = {10.1364/AO.49.004801},
}

@article{Kristensen2023,
  title = {Subnatural Linewidth Superradiant Lasing with Cold $^{88}\mathrm{Sr}$ Atoms},
  author = {Kristensen, Sofus Laguna and Bohr, Eliot and Robinson-Tait, Julian and Zelevinsky, Tanya and Thomsen, Jan W. and M\"uller, J\"org Helge},
  journal = {Phys. Rev. Lett.},
  volume = {130},
  issue = {22},
  pages = {223402},
  numpages = {5},
  year = {2023},
  month = {May},
  publisher = {American Physical Society},
  doi = {10.1103/PhysRevLett.130.223402},
}

@ARTICLE{Rubiola_chart,
  author={Rubiola, Enrico and Vernotte, François},
  journal={IEEE Trans. Microw. Theory Tech.}, 
  title={{The Companion of Enrico’s Chart for Phase Noise and Two-Sample Variances}}, 
  year={2023},
  volume={71},
  number={7},
  pages={2996-3025},
  doi={10.1109/TMTT.2023.3238267}}

@article{LaburtheTolra2026,
  title = {{Instabilities of the Continuous Superradiant Laser}},
  author = {Laburthe-Tolra, Bruno and Robert-de-Saint-Vincent, Martin and Pasquiou, Benjamin},
  journal = {Preprint at arXiv.org},
  year = {2026},
  archivePrefix = {arXiv},
  primaryClass = {quant-ph},
  doi = {10.48550/arXiv.2606.21675},
}

@article{Rugged_mHz_Linewidth_Haonan_2020,
  title = {{Rugged mHz-Linewidth Superradiant Laser Driven by a Hot Atomic Beam}},
  author = {Liu, Haonan and J\"ager, Simon B. and Yu, Xianquan and Touzard, Steven and Shankar, Athreya and Holland, Murray J. and Nicholson, Travis L.},
  journal = {Phys. Rev. Lett.},
  volume = {125},
  issue = {25},
  pages = {253602},
  numpages = {6},
  year = {2020},
  month = {Dec},
  publisher = {American Physical Society},
  doi = {10.1103/PhysRevLett.125.253602},
}

@article{Single-atom_detection_Poldy_2008,
  title = {Single-atom detection with optical cavities},
  author = {Poldy, R. and Buchler, B. C. and Close, J. D.},
  journal = {Phys. Rev. A},
  volume = {78},
  issue = {1},
  pages = {013640},
  numpages = {11},
  year = {2008},
  month = {Jul},
  publisher = {American Physical Society},
  doi = {10.1103/PhysRevA.78.013640},
}

@article{debnathLasingSuperradiantCrossover2018,
  title = {{Lasing in the Superradiant Crossover Regime}},
  author = {Debnath, Kamanasish and Zhang, Yuan and M{\o}lmer, Klaus},
  year = 2018,
  month = dec,
  journal = {Phys. Rev. A},
  volume = {98},
  number = {6},
  pages = {063837},
  issn = {2469-9926, 2469-9934},
  doi = {10.1103/PhysRevA.98.063837}
}

@article{tieriTheoryCrossoverLasing2017,
  title = {{Theory of the Crossover from Lasing to Steady State Superradiance}},
  author = {Tieri, D. A. and Xu, Minghui and Meiser, D. and Cooper, J. and Holland, M. J.},
  journal = {Preprint at arXiv.org},
  year = 2017,
  month = feb,
  number = {arXiv:1702.04830},
  publisher = {arXiv},
  doi = {10.48550/arXiv.1702.04830}
}

@article{confortiAcoustical1Noises2020,
  title = {{Acoustical and} $1/f$ {Noises in Narrow Linewidth Lasers}},
  author = {Conforti, Evandro and Rodigheri, Mareli and Sutili, Tiago and Galdieri, Flavio J.},
  year = 2020,
  month = dec,
  journal = {Opt. Commun.},
  volume = {476},
  pages = {126286},
  issn = {00304018},
  doi = {10.1016/j.optcom.2020.126286}
}

@article{duanGeneralTreatmentThermal2012,
  title = {General Treatment of the Thermal Noises in Optical Fibers},
  author = {Duan, Lingze},
  year = 2012,
  month = aug,
  journal = {Phys. Rev. A},
  volume = {86},
  number = {2},
  pages = {023817},
  issn = {1050-2947, 1094-1622},
  doi = {10.1103/PhysRevA.86.023817},
}

@article{zhangThermalnoiselimitedFiberFrequency2024,
  title = {{A Thermal-Noise-Limited Fiber Frequency Reference with} $0.1\,\mathrm{Hz}/\sqrt{{\rm Hz}}$ {Stability}},
  author = {Zhang, Ya and Bandutunga, Chathura P. and McRae, Terry G. and Gray, Malcolm A. and Chow, Jong H.},
  year = 2024,
  month = nov,
  journal = {J. Phys.: Conf. Ser.},
  volume = {2889},
  number = {1},
  pages = {012052},
  issn = {1742-6588, 1742-6596},
  doi = {10.1088/1742-6596/2889/1/012052}
}

@article{bohnetSteadystateSuperradiantLaser2012a,
  title = {A Steady-State Superradiant Laser with Less than One Intracavity Photon},
  author = {Bohnet, Justin G. and Chen, Zilong and Weiner, Joshua M. and Meiser, Dominic and Holland, Murray J. and Thompson, James K.},
  year = 2012,
  month = apr,
  journal = {Nature},
  volume = {484},
  number = {7392},
  pages = {78--81},
  issn = {0028-0836, 1476-4687},
  doi = {10.1038/nature10920},
  copyright = {http://www.springer.com/tdm}
}

@article{Lee_Frequency_Stability_2026,
  title = {{Frequency Stability of $2.5\ifmmode\times\else\texttimes\fi{}{10}^{\ensuremath{-}17}$ from a Si Cavity with AlGaAs Crystalline Mirrors}},
  author = {Lee, Dahyeon and Hu, Zoey Z. and Lewis, Ben and Aeppli, Alexander and Kim, Kyungtae and Yao, Zhibin and Legero, Thomas and Nicolodi, Daniele and Riehle, Fritz and Sterr, Uwe and Ye, Jun},
  journal = {Phys. Rev. Lett.},
  volume = {136},
  issue = {3},
  pages = {033801},
  numpages = {8},
  year = {2026},
  month = {Jan},
  publisher = {American Physical Society},
  doi = {10.1103/zgrm-cjbb},
}

@article{Aeppli_Atomic_Clock_Frequency_2026,
  title = {Atomic Clock Frequency Ratios with Fractional Uncertainty $\ensuremath{\le}3.2\ifmmode\times\else\texttimes\fi{}{10}^{\ensuremath{-}18}$},
  author = {Aeppli, Alexander and Arthur-Dworschack, Willa J. and Beloy, Kyle and Berry, Caitlin M. and Bothwell, Tobias and Folz, Angela and Fortier, Tara M. and Grogan, Tanner and Hassan, Youssef S. and Hu, Zoey Z. and Hume, David B. and Hunt, Benjamin D. and Kim, Kyungtae and Koepke, Amanda and Lee, Dahyeon and Leibrandt, David R. and Lewis, Ben and Ludlow, Andrew D. and Marshall, Mason C. and Nardelli, Nicholas V. and Ranganath, Harikesh and Castillo, Daniel A. Rodriguez and Sherman, Jeffrey A. and Siegel, Jacob L. and Thornton, Suzanne and Warfield, William and Ye, Jun},
  collaboration = {BACON Collaboration},
  journal = {Phys. Rev. Lett.},
  volume = {137},
  issue = {3},
  pages = {033201},
  numpages = {8},
  year = {2026},
  month = {Jul},
  publisher = {American Physical Society},
  doi = {10.1103/g865-9mk1},
}

@article{Dicke_Coherence_Spontaneous_Radiation_1954,
title = {Coherence in Spontaneous Radiation Processes},
author = {Dicke, R. H.},
journal = {Phys. Rev. },
volume = {93},
issue = {1},
pages = {99--110},
numpages = {12},
year = {1954},
month = {Jan},
publisher = {American Physical Society},
doi = {10.1103/PhysRev.93.99},
}

@article{kuppensQuantumLimitedLinewidthBadCavity1994,
  title = {Quantum-{{Limited Linewidth}} of a {{Bad-Cavity Laser}}},
  author = {Kuppens, S. J. M. and Van Exter, M. P. and Woerdman, J. P.},
  year = 1994,
  month = jun,
  journal = {Phys. Rev. Lett.},
  volume = {72},
  number = {24},
  pages = {3815--3818},
  issn = {0031-9007},
  doi = {10.1103/PhysRevLett.72.3815},
  copyright = {http://link.aps.org/licenses/aps-default-license}
}

@article{vanexterTheoryLinewidthBadcavity1995,
  title = {Theory for the Linewidth of a Bad-Cavity Laser},
  author = {Van Exter, M. P. and Kuppens, S. J. M. and Woerdman, J. P.},
  year = 1995,
  month = jan,
  journal = {Phys. Rev. A},
  volume = {51},
  number = {1},
  pages = {809--816},
  issn = {1050-2947, 1094-1622},
  doi = {10.1103/PhysRevA.51.809},
  copyright = {http://link.aps.org/licenses/aps-default-license}
}

@article{schawlowInfraredOpticalMasers1958,
  title = {Infrared and {{Optical Masers}}},
  author = {Schawlow, A. L. and Townes, C. H.},
  year = 1958,
  month = dec,
  journal = {Phys. Rev.},
  volume = {112},
  number = {6},
  pages = {1940--1949},
  issn = {0031-899X},
  doi = {10.1103/PhysRev.112.1940},
  copyright = {http://link.aps.org/licenses/aps-default-license}
}

@article{arakawaProbingUltralightDark2026,
  title = {Probing {{Ultralight Dark Matter}} at the {{Mega-Planck Scale}} with the {{Thorium Nuclear Clock}}},
  author = {Arakawa, Jason and Doyle, Jack F. and Fuchs, Elina and Higgins, Jacob S. and Kirk, Fiona and Li, Kai and Ooi, Tian and Perez, Gilad and Ratzinger, Wolfram and Safronova, Marianna S. and Schumm, Thorsten and Ye, Jun and Zhang, Chuankun},
  journal = {Preprint at arXiv.org},
  year = {2026},
  archivePrefix = {arXiv},
  primaryClass = {hep-ph},
  doi = {10.48550/arXiv.2602.16804}
}

@article{bothwellResolvingGravitationalRedshift2022,
  title = {Resolving the Gravitational Redshift across a Millimetre-Scale Atomic Sample},
  author = {Bothwell, Tobias and Kennedy, Colin J. and Aeppli, Alexander and Kedar, Dhruv and Robinson, John M. and Oelker, Eric and Staron, Alexander and Ye, Jun},
  year = 2022,
  month = feb,
  journal = {Nature},
  volume = {602},
  number = {7897},
  pages = {420--424},
  issn = {0028-0836, 1476-4687},
  doi = {10.1038/s41586-021-04349-7}
}

@article{chuExploringDynamicalInterplay2025,
  title = {Exploring the {{Dynamical Interplay}} between {{Mass-Energy Equivalence}}, {{Interactions}}, and {{Entanglement}} in an {{Optical Lattice Clock}}},
  author = {Chu, Anjun and {Mart{\'i}nez-Lahuerta}, Victor J. and Miklos, Maya and Kim, Kyungtae and Zoller, Peter and Hammerer, Klemens and Ye, Jun and Rey, Ana Maria},
  year = 2025,
  month = mar,
  journal = {Phys. Rev. Lett.},
  volume = {134},
  number = {9},
  pages = {093201},
  issn = {0031-9007, 1079-7114},
  doi = {10.1103/PhysRevLett.134.093201}
}

@article{kennedyPrecisionMetrologyMeets2020,
  title = {Precision {{Metrology Meets Cosmology}}: {{Improved Constraints}} on {{Ultralight Dark Matter}} from {{Atom-Cavity Frequency Comparisons}}},
  shorttitle = {Precision {{Metrology Meets Cosmology}}},
  author = {Kennedy, Colin J. and Oelker, Eric and Robinson, John M. and Bothwell, Tobias and Kedar, Dhruv and Milner, William R. and Marti, G. Edward and Derevianko, Andrei and Ye, Jun},
  year = 2020,
  month = nov,
  journal = {Phys. Rev. Lett.},
  volume = {125},
  number = {20},
  pages = {201302},
  issn = {0031-9007, 1079-7114},
  doi = {10.1103/PhysRevLett.125.201302}
}

@inproceedings{vernotteThreecorneredHatAllan2016,
  title = {{Three-Cornered Hat versus Allan Covariance}},
  booktitle = {2016 {{IEEE International Frequency Control Symposium}} ({{IFCS}})},
  author = {Vernotte, Francois and Calosso, Claudio Eligio and Rubiola, Enrico},
  year = 2016,
  month = may,
  pages = {1--6},
  publisher = {IEEE},
  address = {New Orleans, LA, USA},
  doi = {10.1109/FCS.2016.7546784},
  isbn = {978-1-5090-2091-1}
}

@article{kazakovUltimateStabilityActive2022,
  title = {Ultimate Stability of Active Optical Frequency Standards},
  author = {Kazakov, Georgy A. and Dubey, Swadheen and Bychek, Anna and Sterr, Uwe and Bober, Marcin and Zawada, Micha{\l}},
  year = 2022,
  month = nov,
  journal = {Phys. Rev. A},
  volume = {106},
  number = {5},
  pages = {053114},
  issn = {2469-9926, 2469-9934},
  doi = {10.1103/PhysRevA.106.053114}
}

@article{abrahamOverviewInstabilitiesLaser1985,
  title = {{Overview of Instabilities in Laser Systems}},
  author = {Abraham, N. B. and Lugiato, L. A. and Narducci, L. M.},
  year = 1985,
  month = jan,
  journal = {J. Opt. Soc. Am.},
  volume = {2},
  number = {1},
  pages = {7},
  issn = {0740-3224, 1520-8540},
  doi = {10.1364/JOSAB.2.000007}
}

@article{ogawaQuasiperiodicInstabilityChaos1988,
  title = {Quasiperiodic Instability and Chaos in the Bad-Cavity Laser with Modulated Inversion: {{Numerical}} Analysis of a {{Toda}} Oscillator System},
  shorttitle = {Quasiperiodic Instability and Chaos in the Bad-Cavity Laser with Modulated Inversion},
  author = {Ogawa, Tetsuo},
  year = 1988,
  month = jun,
  journal = {Phys. Rev. A},
  volume = {37},
  number = {11},
  pages = {4286--4302},
  issn = {0556-2791},
  doi = {10.1103/PhysRevA.37.4286},
  copyright = {http://link.aps.org/licenses/aps-default-license}
}

@article{sundarSqueezingMultilevelAtoms2024,
  title = {Squeezing {{Multilevel Atoms}} in {{Dark States}} via {{Cavity Superradiance}}},
  author = {Sundar, Bhuvanesh and Barberena, Diego and Rey, Ana Maria and Orioli, Asier Pi{\~n}eiro},
  year = 2024,
  month = jan,
  journal = {Phys. Rev. Lett.},
  volume = {132},
  number = {3},
  pages = {033601},
  issn = {0031-9007, 1079-7114},
  doi = {10.1103/PhysRevLett.132.033601}
}

@article{fortierOpticalAtomicClocks2026,
  title = {Optical Atomic Clocks: Defining the Future of Time and Frequency Metrology},
  shorttitle = {Optical Atomic Clocks},
  author = {Fortier, Tara M. and Luiten, Andre N. and Margolis, Helen S.},
  year = 2026,
  month = jan,
  journal = {Optica},
  volume = {13},
  number = {1},
  pages = {143},
  issn = {2334-2536},
  doi = {10.1364/OPTICA.575770}
}

@article{takamotoTestGeneralRelativity2020,
  title = {Test of General Relativity by a Pair of Transportable Optical Lattice Clocks},
  author = {Takamoto, Masao and Ushijima, Ichiro and Ohmae, Noriaki and Yahagi, Toshihiro and Kokado, Kensuke and Shinkai, Hisaaki and Katori, Hidetoshi},
  year = 2020,
  month = jul,
  journal = {Nat. Photonics},
  volume = {14},
  number = {7},
  pages = {411--415},
  issn = {1749-4885, 1749-4893},
  doi = {10.1038/s41566-020-0619-8}
}

@article{parkeLaserFractionalFrequency2026,
  title = {Laser Fractional Frequency Instability at 4 \texttimes 10{\textsuperscript{-17}} with a Room Temperature Optical Reference Cavity},
  author = {Parke, Adam L. and Clulow, Eve and Huang, Wei and Kaur, Namneet and Karembera, Reinhard and Gaudron, Jacques-Olivier and Zhang, Xi and Risaro, Matias and Tunesi, Jacob and Bourne, Henry and Schioppo, Marco},
  year = 2026,
  month = jul,
  journal = {Optica},
  volume = {13},
  number = {7},
  pages = {1190},
  issn = {2334-2536},
  doi = {10.1364/OPTICA.591175}
}

@article{norciaColdStrontiumLaserSuperradiant2016,
  title = {Cold-{{Strontium Laser}} in the {{Superradiant Crossover Regime}}},
  author = {Norcia, Matthew A. and Thompson, James K.},
  year = 2016,
  month = mar,
  journal = {Phys. Rev. X},
  volume = {6},
  number = {1},
  pages = {011025},
  issn = {2160-3308},
  doi = {10.1103/PhysRevX.6.011025},
  copyright = {http://creativecommons.org/licenses/by/3.0/}
}

@article{norciaSuperradianceMillihertzLinewidth2016,
  title = {{Superradiance on the Millihertz Linewidth Strontium Clock Transition}},
  author = {Norcia, Matthew A. and Winchester, Matthew N. and Cline, Julia R. K. and Thompson, James K.},
  year = 2016,
  month = oct,
  journal = {Sci. Adv.},
  volume = {2},
  number = {10},
  pages = {e1601231},
  issn = {2375-2548},
  doi = {10.1126/sciadv.1601231}
}

@article{laburthe-tolraCorrelationsLinewidthAtomic2023,
  title = {Correlations and Linewidth of the Atomic Beam Continuous Superradiant Laser},
  author = {{Laburthe-Tolra}, Bruno and Amodjee, Ziyad and Pasquiou, Benjamin and {Robert-de-Saint-Vincent}, Martin},
  year = 2023,
  month = mar,
  journal = {SciPost Phys. Core},
  volume = {6},
  number = {1},
  pages = {015},
  issn = {2666-9366},
  doi = {10.21468/SciPostPhysCore.6.1.015}
}

@article{dubeyModelingContinuousSuperradiant2025,
  title = {Modeling of a Continuous Superradiant Laser on the Sub-{{mHz}} $^1${{S}}$_0$ {$\rightarrow$} $^3${{P}}$_0$ Transition in Neutral Strontium-88},
  author = {Dubey, Swadheen and Kazakov, Georgy A. and Heizenreder, Benedikt and Zhou, Sheng and Bennetts, Shayne and Sch{\"a}ffer, Stefan Alaric and Sitaram, Ananya and Schreck, Florian and {MoSaiQC Collaboration}},
  year = 2025,
  month = mar,
  journal = {Phys. Rev. Research},
  volume = {7},
  number = {1},
  pages = {013292},
  issn = {2643-1564},
  doi = {10.1103/PhysRevResearch.7.013292}
}

@article{elbadawiFrameworkContinuousSuperradiant2026,
  title = {A Framework for Continuous Superradiant Laser Operation via Sequential Transport of Atoms},
  author = {El Badawi, Jana and Delehaye, Marion and Bellomo, Bruno},
  year = 2026,
  month = apr,
  journal = {C. R. Phys.},
  volume = {27},
  number = {G1},
  pages = {191--216},
  issn = {1631-0705, 1878-1535},
  doi = {10.5802/crphys.277}
}

@article{reillyFullyCollectiveSuperradiant2026a,
  title = {Fully {{Collective Superradiant Lasing}} with {{Vanishing Sensitivity}} to {{Cavity Length Vibrations}}},
  author = {Reilly, Jarrod T. and J{\"a}ger, Simon B. and Cooper, John and Holland, Murray J.},
  year = 2026,
  month = apr,
  journal = {Phys. Rev. Lett.},
  volume = {136},
  number = {14},
  pages = {143803},
  issn = {0031-9007, 1079-7114},
  doi = {10.1103/v6jq-m6sk}
}

@article{tangProspectsSuperradiantLaser2022,
  title = {Prospects of a Superradiant Laser Based on a Thermal or Guided Beam of {{Sr}} 88},
  author = {Tang, Mikkel and Sch{\"a}ffer, Stefan A. and M{\"u}ller, J{\"o}rg H.},
  year = 2022,
  month = dec,
  journal = {Phys. Rev. A},
  volume = {106},
  number = {6},
  pages = {063704},
  issn = {2469-9926, 2469-9934},
  doi = {10.1103/PhysRevA.106.063704}
}

@article{zhangActiveFrequencyMeasurement2022,
  title = {Active {{Frequency Measurement}} on {{Superradiant Strontium Clock Transitions}}},
  author = {Zhang, Yuan and Shan, Chongxin and M{\o}lmer, Klaus},
  year = 2022,
  month = jan,
  journal = {Phys. Rev. Lett.},
  volume = {128},
  number = {1},
  pages = {013604},
  issn = {0031-9007, 1079-7114},
  doi = {10.1103/PhysRevLett.128.013604}
}

@article{famaCollimationDenseAtomic2026,
  title = {Collimation of Dense Atomic Beams by {{Swept Velocity Shelving}}},
  author = {Fam{\`a}, Francesca and Heizenreder, Benedikt and Sitaram, Ananya and Beli, Camila and Spreeuw, Robert J. C. and Sch{\"a}ffer, Stefan A. and Schreck, Florian},
  journal = {Preprint at arXiv.org},
  year = 2026,
  month = aug,
  publisher = {arXiv},
  doi = {10.48550/ARXIV.2607.28873}
}

@article{famaContinuousCavityQED2024,
  title = {Continuous Cavity {{QED}} with an Atomic Beam},
  author = {Fam{\`a}, Francesca and Zhou, Sheng and Heizenreder, Benedikt and Tang, Mikkel and Bennetts, Shayne and J{\"a}ger, Simon B. and Sch{\"a}ffer, Stefan A. and Schreck, Florian},
  year = 2024,
  month = dec,
  journal = {Phys. Rev. A},
  volume = {110},
  number = {6},
  pages = {063721},
  issn = {2469-9926, 2469-9934},
  doi = {10.1103/PhysRevA.110.063721}
}

@article{kirtonIntroductionDickeModel2019,
  title = {Introduction to the {{Dicke Model}}: {{From Equilibrium}} to {{Nonequilibrium}}, and {{{\emph{Vice Versa}}}}},
  shorttitle = {Introduction to the {{Dicke Model}}},
  author = {Kirton, Peter and Roses, Mor M. and Keeling, Jonathan and Dalla Torre, Emanuele G.},
  year = 2019,
  month = feb,
  journal = {Adv. Quantum Technol.},
  volume = {2},
  number = {1-2},
  pages = {1800043},
  issn = {2511-9044, 2511-9044},
  doi = {10.1002/qute.201800043}
}

@article{mivehvarDrivenDissipativeSupersolidRing2018,
  title = {Driven-{{Dissipative Supersolid}} in a {{Ring Cavity}}},
  author = {Mivehvar, Farokh and Ostermann, Stefan and Piazza, Francesco and Ritsch, Helmut},
  year = 2018,
  month = mar,
  journal = {Phys. Rev. Lett.},
  volume = {120},
  number = {12},
  pages = {123601},
  issn = {0031-9007, 1079-7114},
  doi = {10.1103/PhysRevLett.120.123601}
}

@article{yi-xiangGoldstoneHiggsModes2013,
  title = {Goldstone and {{Higgs}} Modes of Photons inside a Cavity},
  author = {{Yi-Xiang}, Yu and Ye, Jinwu and Liu, Wu-Ming},
  year = 2013,
  month = dec,
  journal = {Sci. Rep.},
  volume = {3},
  number = {1},
  pages = {3476},
  issn = {2045-2322},
  doi = {10.1038/srep03476}
}

@article{zhangExtremelyBadcavityLaser2024,
  title = {{An Extremely Bad-Cavity Laser}},
  author = {Zhang, Jia and Shi, Tiantian and Miao, Jianxiang and Yu, Deshui and Chen, Jingbiao},
  year = 2024,
  month = sep,
  journal = {npj Quantum Inf.},
  volume = {10},
  number = {1},
  pages = {87},
  issn = {2056-6387},
  doi = {10.1038/s41534-024-00880-3}
}

@article{liuZeroDeadTimeStrontiumLattice2025,
  title = {Zero-{{Dead-Time Strontium Lattice Clock}} with a {{Stability}} at $10^{-19}$ {{Level}}},
  author = {Liu, Xiao-Yong and Liu, Peng and Li, Jie and Zhang, Yu-Chen and Wang, Yuan-Bo and Jia, Zhi-Peng and Zhang, Xiang and Zhu, Xian-Qing and Kong, De-Quan and Song, Wen-Lan and Niu, Guo-Zhen and Yang, Yu-Meng and Feng, Pei-Jun and Liu, Xiang-Pei and Cui, Xing-Yang and Xu, Ping and Jiang, Xiao and Yin, Juan and Liao, Sheng-Kai and Peng, Cheng-Zhi and Dai, Han-Ning and Chen, Yu-Ao and Pan, Jian-Wei},
  year = 2025,
  month = dec,
  journal = {Phys. Rev. Lett.},
  volume = {135},
  number = {26},
  pages = {263402},
  issn = {0031-9007, 1079-7114},
  doi = {10.1103/zbpb-6qxb}
}

@article{daiReviewDevelopmentHydrogen2022,
  title = {{Review of the Development of the Hydrogen Maser Technique and a Brief Introduction to Its Space Applications}},
  author = {Dai, Jiayu and Liu, Tiexin and Cai, Yong and Chen, Zhichun and Li, Qi},
  year = 2022,
  month = sep,
  journal = {Front. Phys.},
  volume = {10},
  pages = {995361},
  issn = {2296-424X},
  doi = {10.3389/fphy.2022.995361}
}

@article{leibrandtSphericalReferenceCavities2011,
  title = {Spherical Reference Cavities for Frequency Stabilization of Lasers in Non-Laboratory Environments},
  author = {Leibrandt, David R. and Thorpe, Michael J. and Notcutt, Mark and Drullinger, Robert E. and Rosenband, Till and Bergquist, James C.},
  year = 2011,
  month = feb,
  journal = {Opt. Express},
  volume = {19},
  number = {4},
  pages = {3471},
  issn = {1094-4087},
  doi = {10.1364/OE.19.003471},
  copyright = {https://doi.org/10.1364/OA\_License\_v1\#VOR-OA}
}

@article{Zhou_Laser_Frequency_Shift_2020,
title = {{Laser frequency shift up to 5 GHz with a high-efficiency 12-pass 350-MHz acousto-optic modulator}},
author = {Zhou, Chao and He, Chuan and Yan, Si-Tong and Ji, Yu-Hang and Zhou, Lin and Wang, Jin and Zhan, Ming-Sheng},
journal = {Rev. Sci. Instrum.},
volume = {91},
issue = {3},
pages = {033201},
numpages = {7},
year = {2020},
month = {Mar},
publisher = {AIP Publishing},
doi = {10.1063/1.5142314}
}

@phdthesis{urech2023single,
  author       = {A. A. Urech},
  title        = {Single strontium atoms in optical tweezers},
  school       = {University of Amsterdam},
  year         = {2023},
  url          = {https://dare.uva.nl/id/cb589140-c918-4720-a40a-9b82071c1fa1}
}

@article{ferioliNonequilibriumSuperradiantPhase2023,
  title = {A Non-Equilibrium Superradiant Phase Transition in Free Space},
  author = {Ferioli, Giovanni and Glicenstein, Antoine and {Ferrier-Barbut}, Igor and Browaeys, Antoine},
  year = 2023,
  month = sep,
  journal = {Nat. Phys.},
  volume = {19},
  number = {9},
  pages = {1345--1349},
  issn = {1745-2473, 1745-2481},
  doi = {10.1038/s41567-023-02064-w}
}

@article{kimCoherentSingleatomSuperradiance2018,
  title = {Coherent Single-Atom Superradiance},
  author = {Kim, Junki and Yang, Daeho and Oh, Seung-hoon and An, Kyungwon},
  year = 2018,
  month = feb,
  journal = {Science},
  volume = {359},
  number = {6376},
  pages = {662--666},
  issn = {0036-8075, 1095-9203},
  doi = {10.1126/science.aar2179}
}

@article{ozturkObservationNonHermitianPhase2021,
  title = {Observation of a Non-{{Hermitian}} Phase Transition in an Optical Quantum Gas},
  author = {{\"O}zt{\"u}rk, Fahri Emre and Lappe, Tim and Hellmann, G{\"o}ran and Schmitt, Julian and Klaers, Jan and Vewinger, Frank and Kroha, Johann and Weitz, Martin},
  year = 2021,
  month = apr,
  journal = {Science},
  volume = {372},
  number = {6537},
  pages = {88--91},
  issn = {0036-8075, 1095-9203},
  doi = {10.1126/science.abe9869}
}

@article{Riley2008,
  author       = {W.J. Riley},
  title        = {Handbook of Frequency Stability Analysis},
  journal = {NIST Special Publication 1065},
  year         = {2008},
  month        = jul,
  publisher    = {National Institute of Standards and Technology},
  address      = {Boulder, CO},
  series       = {NIST Special Publication 1065},
}

@article{Arnold2026,
  author    = {Arnold, K. J. and Lee, M. D. K. and Zhao, Qi and Qin, Qichen and Zhang, Zhao and Jayjong, N. and Barrett, M. D.},
  title     = {{Lu$^+$} Optical Frequency References with Accuracy Verified at the 19th Digit},
  journal   = {Nature},
  year      = {2026},
  month     = oct,
  volume    = {658},
  number    = {8134},
  pages     = {60--64},
  doi       = {10.1038/s41586-026-11072-8},
  issn      = {1476-4687}
}

@article{Mehlstaubler2018,
  author    = {Mehlst{\"a}ubler, Tanja E. and Grosche, Gesine and Lisdat, Christian and Schmidt, Piet O. and Denker, Heiner},
  title     = {Atomic Clocks for Geodesy},
  journal   = {Rep. Prog. Phys.},
  year      = {2018},
  month     = apr,
  volume    = {81},
  number    = {6},
  pages     = {064401},
  publisher = {IOP Publishing},
  doi       = {10.1088/1361-6633/aab409}
}

@INPROCEEDINGS{Parker_hydrogen_masers_2010,
  author={Parker, T.E. and Jefferts, S.R. and Heavner, T.P.},
  booktitle={2010 IEEE International Frequency Control Symposium}, 
  title={Medium-term frequency stability of hydrogen masers as measured by a cesium fountain}, 
  year={2010},
  volume={},
  number={},
  pages={318-323},
  doi={10.1109/FREQ.2010.5556319}}

@article{Continuous_Collective_Cline_2025,
  title = {{Continuous Collective Strong Coupling of Strontium Atoms to a High Finesse Ring Cavity}},
  author = {Cline, Julia R. K. and Sch{\"a}fer, Vera M. and Niu, Zhijing and Young, Dylan J. and Yoon, Tai Hyun and Thompson, James K.},
  journal = {Phys. Rev. Lett.},
  volume = {134},
  issue = {1},
  pages = {013403},
  numpages = {6},
  year = {2025},
  month = {Jan},
  publisher = {American Physical Society},
  doi = {10.1103/PhysRevLett.134.013403}
}

@misc{Cont_SR_data,
  author       = {B. Heizenreder and S. B. Jäger and F. Famà and C. Beli Silva and Y. Cui and S. A. Schäffer and A. Sitaram and F. Schreck},
  title        = {Continuous superradiant lasing on a narrow optical transition -- Data repository},
  year         = {2026},
  doi          = {10.21942/uva.35025481},
  note         = {Data repository}
}
\end{document}